\documentclass[fleqn,usenatbib]{mnras}

\usepackage{newtxtext,newtxmath}
\usepackage[T1]{fontenc}

\usepackage{graphicx}	% Including figure files
\usepackage{amsmath}	% Advanced maths commands
\usepackage{xcolor}
\usepackage{soul}

\definecolor{bluehl}{rgb}{0.75,0.75,1}

\newcommand{\HI}{H{\,\sc{i}}\ }

\title[\HI maps in TNG50]{The atomic-to-molecular hydrogen transition in the TNG50 simulation: Using realistic UV fields to create spatially resolved \HI maps}

\author[A. Gebek et al.]{%
Andrea Gebek,$^1$\thanks{E-mail: andrea.gebek@ugent.be}
Maarten Baes,$^1$
Benedikt Diemer,$^2$
W.J.G. de Blok,$^{3,4,5}$
Dylan Nelson,$^6$
\newauthor
Anand Utsav Kapoor, $^1$
Peter Camps, $^1$
Omphile Rabyang, $^7$
and Lerothodi Leeuw $^7$
\\
$^1$Sterrenkundig Observatorium, Universiteit Gent, Krijgslaan 281 S9, 9000 Gent, Belgium\\
$^2$Department of Astronomy, University of Maryland, College Park, MD 20742, USA\\
$^3$Netherlands Institute for Radio Astronomy (ASTRON), Oude Hoogeveensedijk 4, 7991 PD, Dwingeloo, Netherlands\\
$^4$Kapteyn Astronomical Institute, PO Box 800, 9700 AV Groningen, Netherlands\\
$^5$Department of Astronomy, University of Cape Town, Private Bag X3, Rondebosch 7701, South Africa\\
$^6$Institut für Theoretische Astrophysik, Zentrum für Astronomie, Universität Heidelberg, Albert-Ueberle-Str. 2, 69120 Heidelberg, Germany\\
$^7$Department of Physics and Astronomy, University of the Western Cape, Private Bag X17, Bellville 7535, South Africa
}

\pubyear{2022}

\begin{document}
\label{firstpage}
\pagerange{\pageref{firstpage}--\pageref{lastpage}}
\maketitle

\begin{abstract}
Cold gas in galaxies provides a crucial test to evaluate the realism of cosmological hydrodynamical simulations. To extract the atomic and molecular hydrogen properties of the simulated galaxy population, postprocessing methods taking the local UV field into account are required. We improve upon previous studies by calculating realistic UV fields with the dust radiative transfer code SKIRT to model the atomic-to-molecular transition in TNG50, the highest-resolution run of the IllustrisTNG suite. Comparing integrated quantities such as the \HI mass function, we study to what detail the UV field needs to be modelled in order to calculate realistic cold gas properties. We then evaluate new, spatially resolved comparisons for cold gas in galaxies by exploring synthetic maps of atomic hydrogen at redshift zero and compare them to 21-cm observations of local galaxies from the WHISP survey. In terms of non-parametric morphologies, we find that TNG50 \HI maps are less concentrated than their WHISP counterparts (median $\Delta C\approx0.3$), due in part to central \HI deficits related to the ejective character of supermassive black hole feedback in TNG. In terms of the \HI column density distribution function, we find discrepancies between WHISP and IllustrisTNG that depend on the total \HI abundance in these datasets as well as the postprocessing method. To fully exploit the synergy between cosmological simulations and upcoming deep \HI{}/H$_2$ data, we advocate the use of accurate methods to estimate the UV radiation field and to generate mock maps.

\end{abstract}


%%%%%%%%%%%%%%%%%%%%%%%%%%%%%%%%%%%%%%%%%%%%%%%%%%

%%%%%%%%%%%%%%%%% BODY OF PAPER %%%%%%%%%%%%%%%%%%

\section{Introduction}

The physical processes regulating the interstellar medium (ISM) are integral to understand the evolution of galaxies. The most important process in galaxy evolution, star formation, occurs in the cold and dense molecular ISM. For molecular hydrogen to form, enough atomic gas (potentially supported by the presence of dust) is required to cool and shield the molecular clouds against energetic ultraviolet (UV) radiation. This leads to the empirical correlation between the surface densities of gas and star-formation rate (the Kennicutt-Schmidt law, \citealt{Schmidt1959}; \citealt{Kennicutt1998}), which is significantly tighter when using the surface density of molecular gas on sub-kpc scales (e.g. \citealt{Bigiel2008}). On the other hand, the ISM is actively shaped by various processes of galaxy evolution such as momentum and energy injections from supernovae and active galactic nuclei, chemical evolution due to metal return from evolved stars, large-scale galactic inflows from the gas residing in the halo, and interactions with other galaxies like ram-pressure stripping. These processes shape the ISM into a complex, multi-phase environment which is a prime target for observational campaigns targeted at understanding galaxy evolution.

As a dominant fraction of the mass of the ISM exists in the form of atomic hydrogen (\HI), observations targeted at the 21-cm line of \HI are useful to study the structure of nearby galaxies. Unlike Ly-$\alpha$ radiation, photons emitted due to the \HI spin-flip transition which gives rise to the 21-cm line penetrate both interstellar dust clouds and Earth's atmosphere. \HI exists in the ISM mostly in two equilibrium states, the warm ($T\sim5000\,$K) and the cold ($T\sim100\,$K) neutral media (WNM and CNM, e.g. \citealt{Saintonge2022}). 21-cm emission from the WNM is generally assumed to be optically thin which gives a trivial relation between the observable 21-cm flux density and the physical quantity of interest, the \HI mass surface density. 21-cm observations of the Milky Way indicate that the WNM contains most of the \HI mass (\citealt{Murray2018}). The CNM is dense and cold enough for \HI self-absorption to become important which complicates the relation between 21-cm flux and \HI mass. For 21-cm emission observations, correction factors to account for the `hidden' \HI mass in the CNM are rather uncertain. While earlier measurements in the LMC, M31 and M33 indicated a correction factor of $\approx35\,\%$ (\citealt{Braun2012}), the more recent study by \citet{Koch2021} finds for the same objects a correction factor of $\approx10\,\%$, in line with the ALFALFA survey of local galaxies (\citealt{Jones2018}).

Observational campaigns using 21-cm observations to measure the galactic \HI content of local galaxies include blind H{\,\sc{i}}-selected surveys (HIPASS, \citealt{Barnes2001}; ALFALFA, \citealt{Giovanelli2005}) and surveys targeting samples representative of the galaxy population (GASS, \citealt{Catinella2010}; xGASS, \citealt{Catinella2018}). These spatially unresolved surveys made use of the superior sensitivity of the single-dish Parkes and Arecibo telescopes to measure galactic \HI masses. With these data, multiple scaling relations between the atomic gas mass and other galactic properties such as stellar mass and star formation rate have been established (see \citealt{Saintonge2022} for a review), with far-reaching implications for galaxy evolution. Interferometric observations obtained spatially resolved 21-cm fluxes to map the structure and kinematics of \HI (e.g. WHISP, \citealt{Hulst2001}; THINGS, \citealt{Walter2008}; BLUEDISK, \citealt{Wang2013}). Such resolved data has been used to study radial \HI surface density profiles (\citealt{Wang2014}), \HI morphologies in the context of mergers (\citealt{Holwerda2011II}), and the Tully-Fisher relation (\citealt{Tully1977}) constructed from spatially resolved \HI kinematics (\citealt{Ponomareva2017}).

These observational insights into the cold gas content of galaxies provide a vital test for simulations of galaxy evolution. Modelling atomic and molecular hydrogen within galaxies numerically requires resolving the molecular clouds in the CNM to capture the formation of H$_2$ within a chemical network. For cosmological hydrodynamical simulations that follow the evolution of baryons and dark matter of large volumes ($\sim 100^3\,\mathrm{cMpc^3}$), modelling this chemical network on the required spatial scales is not feasible at present. Moreover, this chemical network depends on the local radiation field, which for current large-volume cosmological simulations run to $z=0$ is not explicitly followed. Hence, as a critical test for cosmological simulations, postprocessing of the simulation output is required to extract the atomic and molecular hydrogen content and compare to existing or future observational data. Such postprocessing studies (EAGLE: \citealt{Lagos2015}; \citealt{Bahe2016}; \citealt{Marasco2016}; \citealt{Crain2017}; AURIGA: \citealt{Marinacci2017}; IllustrisTNG: \citealt{Diemer2018}; \citealt{Villaescusa2018}; \citealt{Popping2019}; \citealt{Diemer2019}; \citealt{Stevens2019}; \citealt{Watts2020}; \citealt{Inoue2020}; \citealt{Stevens2021}; \citealt{Yates2021}; SIMBA: \citealt{Dave2020}; FireBOX: \citealt{Gensior2022}) mostly focus on redshift zero, where observational data is most available. Despite qualitative agreement, inferred discrepancies include the presence of spurious \HI holes in the EAGLE simulation (\citealt{Bahe2016}), an excess of \HI and too large \HI disks in the AURIGA simulation (\citealt{Marinacci2017}) and an overabundance of \HI at $z=0$ for TNG50 (\citealt{Diemer2019}).  

For molecular hydrogen, a robust comparison between simulations and observations is much more challenging (e.g. \citealt{Popping2019}; \citealt{Inoue2020}). In contrast to the direct detectability of H{\,\sc{i}}, molecular hydrogen is very difficult to observe directly due to the lack of a permanent dipole moment and high excitation temperatures. The molecular gas is usually observed through transitions of the second most abundant molecule, CO, introducing a rather uncertain factor to convert the CO line luminosity to the H$_2$ mass (see \citealt{Bolatto2013} for a review).

Key to H{\,\sc{i}}/H$_2$ postprocessing studies is a scheme to partition the neutral hydrogen simulation output into its atomic and molecular phases. Partitioning schemes based on analytical arguments or high-resolution galaxy simulations both consider the formation of H$_2$ on dust grains and H$_2$ photodissociation from UV radiation in the Lyman-Werner band ($11.2-13.6\,\mathrm{eV}$, $912-1108\,$\AA). Hence, applying such hydrogen partitioning models to cosmological simulations requires estimating the local radiation field strength at $1000\,$\AA\ since the partitioning models use this value as a proxy for the radiation field strength in the Lyman-Werner band. Two different approximations to estimate the UV field exist in the literature: \citet{Lagos2015} introduced a scaling with the local star-formation rate of gas cells. Cosmological simulations typically use a Kennicutt-Schmidt type relation above a specific density threshold to model star formation, while lower-density cells are not star-forming (see Section~\ref{sec:IllustrisTNG}). This leads to a strong discontinuity in the estimated UV field. To overcome this limitation, \citet{Diemer2018} introduced a method which spreads the UV flux from star-forming gas cells without attenuation (but with an escape fraction of 10\%) throughout the galaxy. Since attenuation by dust is significant in the UV these radiation field estimates introduce large modelling uncertainties. 

In this study we improve upon previous methods by using the Monte Carlo dust radiative transfer code SKIRT (\citealt{Baes2011}; \citealt{Camps2015}; \citealt{Camps2020}) to obtain a realistic estimate for the UV field. We describe our simulation methods and the simulation and observational datasets in Section~\ref{sec:Methods}. Using the highest-resolution installment of the IllustrisTNG simulation suite, TNG50-1, we explore how the different UV field estimates affect the H{\,\sc{i}}/H$_2$ statistics of the simulated galaxy population in terms of mass functions and average radial profiles in Section~\ref{sec:UV comparison}. Exploiting the realistic UV fields and the high resolution of TNG50-1, we evaluate the realism of cold gas in cosmological simulations by generating spatially resolved \HI maps. We compare the simulated \HI maps to 21-cm data from the WHISP (Westerbork \HI survey of Spiral and Irregular Galaxies) survey in terms of their non-parametric morphologies in Section~\ref{sec:morphologies}. As complementary metric to non-parametric morphologies, we consider \HI column density distribution functions (tracing \textit{how much} \HI exists per column density instead of \text{how} the \HI columns are spatially distributed) in Section~\ref{sec:HI CDDF}. We discuss and contextualize our results in Section~\ref{sec:Discussion}, and conclude in Section~\ref{sec:Conclusions}.

\section{Methods}\label{sec:Methods}

\subsection{The IllustrisTNG simulations}\label{sec:IllustrisTNG}

The IllustrisTNG suite (\citealt{TNG_Pillepich}; \citealt{TNG_Springel}; \citealt{TNG_Nelson}; \citealt{TNG_Naiman}; \citealt{TNG_Marinacci}) is a set of cosmological, magnetohydrodynamical simulations run using the moving-mesh code AREPO (\citealt{Springel2010}). The simulation suite consists of three different volumes with box sizes of approximately 50, 100, and 300 comoving Mpc, each realized with three to four different resolutions. All of these simulations were run with the same physical model, which means that their subgrid parameters were not recalibrated (unlike the EAGLE simulation, \citealt{Schaye2015}). For the cosmological parameters, the simulations use the 2015 results measured by the Planck satellite (\citealt{Planck2016}). Since we explore synthetic \HI maps which are sensitive to the spatial resolution we use the simulation with the highest resolution, TNG50-1, hereafter referred to as TNG50 (\citealt{Pillepich2019}; \citealt{Nelson2019}). We also consider the lower-resolution TNG50-2 and the larger-volume TNG100-1 (hereafter referred to as TNG100) runs to test the convergence of some of our results against resolution and box size. We summarize the box sizes and resolutions of the different IllustrisTNG simulations considered in this study in Table~\ref{tab:TNGruns}. In the following, we briefly describe the aspects of IllustrisTNG and its galaxy formation model (\citealt{Weinberger2017}; \citealt{Pillepich2018}) that are most relevant to this study.

TNG50 simulates a cube with box size of 51.7 comoving Mpc from $z=127$ to $z=0$. This volume is resolved with $2160^3$ baryonic and dark matter particles, corresponding to a mean particle mass of $8.5\times10^4\,M_\odot$ and $4.5\times10^5\,M_\odot$, respectively. This mass resolution enables a spatial resolution of $70-140\,\mathrm{pc}$ for the densest star-forming regions of galaxies. Galaxies are identified using the SUBFIND algorithm (\citealt{Springel2001}). The IllustrisTNG model incorporates gas radiative processes including metal-line cooling, evolution of stellar populations and chemical enrichment, and feedback from supernovae and black holes that drives galactic outflows and winds. Since molecular clouds cannot be resolved in the simulation, star formation is modelled stochastically for gas with $n_\mathrm{H}>0.106\,\mathrm{cm}^{-3}$ according to the two-phase model of \citet{Springel2003}. The ISM above this density threshold is emulated as cold, star-forming clouds with $T=1000\,\mathrm{K}$ embedded in hot, ionized gas. This model prescribes an effective equation of state by calculating effective pressures and temperatures as averages over the cold clouds and the hot gas.

As we are interested in the cold gas properties of local galaxies we select all subhalos in TNG50 at $z=0$ with a gas mass larger than $10^7\,M_\odot$, such that the galaxies are resolved by at least $\approx100$ gas cells. Furthermore, the SKIRT calculation requires star particles to estimate the radiation field. We choose a minimum total stellar mass of $10^7\,M_\odot$ such that galaxies are resolved by at least $\approx100$ stellar particles. These criteria lead to a sample of 12'431 galaxies, comprising both centrals and satellites. This provides a broad base sample which contains the vast majority of galaxies detectable in 21-cm observations. To facilitate the comparison between the different simulation runs we apply the same galaxy sample criteria ($M_\mathrm{gas}>10^7\,\mathrm{M}_\odot$, $M_\star>10^7\,\mathrm{M}_\odot$) to TNG50-2 and TNG100 as well (the sizes of the galaxy samples are given in Table~\ref{tab:TNGruns}).

For all calculations in this study we consider the subhalos in the simulation individually, meaning that when modelling the H{\,\sc{i}}/H$_2$ content of a galaxy we only select gas cells and star particles that are bound to this specific galaxy as identified by the SUBFIND algorithm. \citet{Villaescusa2018} find that for TNG100 at $z=0$, $\approx98\,\%$ of the \HI gas mass is bound to subhalos (for H$_2$ we expect an even higher fraction), hence we do not miss substantial amounts of H{\,\sc{i}}/H$_2$ gas with our methodology.

\begin{table}
    \centering
    \begin{tabular}{cccc}
         Simulation & $V_\mathrm{sim}\,[\mathrm{cMpc}^3]$ & $m_\mathrm{baryon}\,[10^4\,\mathrm{M}_\odot]$ & $N_\mathrm{gal}$ \\ \hline
         TNG50-1 & $51.7^3$ & 8.5 & 12'431\\
         TNG50-2 & $51.7^3$ & 68 & 9'044\\
         TNG100-1 & $106.5^3$ & 140 & 74'460
         \end{tabular}
    \caption{Runs of the IllustrisTNG suite that we consider in this study. For each simulation, we list the volume, the target baryon mass (the resolution), and the number of galaxies that conform to our sample selection criteria.}
    \label{tab:TNGruns}
\end{table}

\subsection{UV field estimates}\label{sec:U_MW}

\begin{table}
    \centering
    \begin{tabular}{ccc}
         UV field scheme & UV source & Radiation distribution\\ \hline
         Lagos$^1$ & Star-forming gas cells & None\\
         Diemer$^2$ & Star-forming gas cells & Optically thin\\
         SKIRT$^3$ & Star particles & Dust radiative transfer
         \end{tabular}
    \caption{Overview of the different UV field schemes considered in this work. $^1$\citet{Lagos2015}, $^2$\citet{Diemer2018}, $^3$this work.}
    \label{tab:UVfields}
\end{table}

Radiation in the Lyman-Werner band ($912-1108\,$\AA) can photodissociate molecular hydrogen. Consequently, the radiation field strength at 1000\,\AA{} (generally used as a proxy for the radiation of the entire Lyman-Werner band) is a key parameter for many partitioning schemes. We use three different methods to calculate the $U_\mathrm{MW}$ parameter, which is the UV field at 1000\,\AA{} normalized to the average Milky Way value of $3.43\times10^{-8}\,\mathrm{photons\,s^{-1}cm^{-2}Hz^{-1}}$ (\citealt{Draine1978}). All methods rely on estimating the flux from stars and/or star-forming regions. For all gas cells we set a floor on $U_\mathrm{MW}$ of 0.00137 which corresponds to the homogeneous UV background (UVB) of \citet{Faucher2009} (in the updated 2011 version) at $z=0$ (the same UVB model is implemented in IllustrisTNG for the calculation of the ionization state of the gas). Dense gas could be self-shielded against the UVB which would lower the floor value of $U_\mathrm{MW}$. However, such gas cells are typically near high star-formation areas so that the local UV radiation surpasses the UVB, rendering the actual floor value irrelevant. We consider three different schemes to estimate the UV field, summarized in Table~\ref{tab:UVfields}.

The simplest method to estimate the UV field follows \citet{Lagos2015}. The `Lagos' approximation is based on the insight that the largest fraction of the UV flux typically comes from very young stars. $U_\mathrm{MW}$ is calculated by scaling the local star formation rate surface density by the typical Milky Way value:

\begin{equation}
    U_\mathrm{MW}=\frac{\mathrm{SFR}\cdot\rho/m\cdot\lambda_\mathrm{J}}{ 10^{-9}\,M_\odot\,\mathrm{pc}^{-2}},
\end{equation}
where $\rho$ is the total gas mass density, $m$ the gas cell mass, SFR its star-formation rate, and $\lambda_\mathrm{J}$ the Jeans length which approximates the size of a self-gravitating gas cloud (\citealt{Schaye2001}; \citealt{Schaye2008}):

\begin{equation}
    \lambda_\mathrm{J}=\sqrt{\frac{\gamma(\gamma-1)u}{G\rho}}.
\end{equation}
In this equation, $\gamma=5/3$ is the ratio of specific heat capacities, $u$ the internal energy per unit mass, and $G$ the gravitational constant. Due to temporal and mass resolution limits in the simulation, it is impossible to resolve arbitrarily small star-formation rates for the gas cells. This leads to a minimum value for $U_\mathrm{MW}$ in the Lagos approach for the star-forming gas cells on the order of unity, while all gas cells with zero star-formation rate have $U_\mathrm{MW}=0.00137$ (the UV background). Hence, this UV field approximation creates a substantial, unphysical dichotomy in the UV field distribution of the gas cells.

\citet{Diemer2018} introduce an improved UV field calculation to overcome this limitation. In the `Diemer' approximation, star-forming gas cells are assigned a UV flux based on a Starburst99 (\citealt{Leitherer1999}) calculation of a continuously forming population of stars following a Kroupa IMF (\citealt{Kroupa2001}). This UV flux is scaled to the star-formation rate of the gas cell, the flux of a $1\,M_\odot\mathrm{yr}^{-1}$ cell at a distance of $1\,\mathrm{kpc}$ is $3.3\times10^{-6}\,\mathrm{photons\,s^{-1}cm^{-2}Hz^{-1}}$ corresponding to $U_\mathrm{MW}=96.2$. It is assumed that a certain fraction of this radiation is absorbed by dust within the star-forming region and the remaining fraction propagates through a transparent medium. \citet{Diemer2018} calibrated this escape fraction to $10\,\%$ based on the SFR-UV relation in the solar neighbourhood. The propagation of the UV flux is calculated via a Fourier transform on a regular cubic grid. We refer the reader to Appendix A of \citet{Diemer2018} for details.

While the `Diemer' method models in-situ absorption of UV flux generated from star-forming regions, it does not capture the significant absorption by diffuse dust in the ISM. Furthermore, older stellar populations can also contribute significantly to the galactic UV flux (\citealt{Viaene2016}; \citealt{Bianchi2018}; \citealt{Nersesian2019}). To generate realistic UV fields taking the complex galactic star-dust geometries into account, 3D radiative transfer modelling is required.

Here we use the Monte Carlo dust radiative transfer code \mbox{SKIRT 9} (\citealt{Camps2015}; \citealt{Camps2020}). We run SKIRT in the oligochromatic (i.e. single wavelength) mode at 1000\,\AA \, including dust attenuation. A SKIRT simulation requires to define a box to set the domain for the radiative transfer calculation. We use a cube with side length $2\cdot r_\mathrm{A}$\footnote{The same cube size is used to compute the Diemer UV field.}, with the aperture radius $r_\mathrm{A}$ chosen such that a sphere with this radius contains at least 99.9\,\% of the neutral hydrogen gas mass bound to a specific subhalo (see Section~\ref{sec:HI fraction} for the calculation of the neutral hydrogen fraction). We choose the different components in a SKIRT simulation following \citet{Kapoor2021}; \citet{Trcka2022} as follows:

\begin{itemize}

\item Evolved stellar populations: We select star particles bound to the subhalo within the spherical aperture to model the UV emission. Star particles with ages above 10\,Myr are treated as evolved stellar populations and modelled with a Bruzual-Charlot spectral energy distribution (\citealt{Bruzual2003}) with Chabrier initial mass function (\citealt{Chabrier2003}). We have verified that the choice of an alternative stellar population library (BPASS, \citealt{Eldridge2017}; \citealt{Stanway2018}) does not affect our results.

\item Star-forming regions: Bound star particles within the spherical aperture and with ages below 10\,Myr are treated as star-forming regions and modelled with Mappings-III templates (\citealt{Groves2008}). These templates contain the emission of the young stellar population and its subsequent attenuation by the surrounding dusty birth cloud. We refer to \citet{Trcka2022} for the determination of the various parameters required for the Mappings-III templates such as the compactness of the star-forming region or the ISM pressure.

\item Diffuse dust: To track the radiation field and its attenuation by diffuse dust we select all gas cells bound to the subhalo within a cube of side length $2\cdot r_\mathrm{A}$. We allocate dust only in relatively dense and cold gas cells that are in line with the criterion from \citet{Torrey2012}. We assume that $40\,\%$ of the metallic mass in these gas cells exists as dust (\citealt{Liang2023} and references therein), which we model with the THEMIS dust mix (\citealt{Jones2017}). 

\end{itemize}

The spatial discretization within SKIRT uses the imported Voronoi cell centers of the gas cells (\citealt{Camps2013}). This allows SKIRT to reconstruct the same Voronoi mesh that was used in IllustrisTNG. Tracking the photon packages through a Voronoi mesh is computationally more expensive than octree grids, but the advantage of importing a grid instead of calculating one outweighs the additional cost here (see also \citealt{Camps2015}). We find converged results when running SKIRT with $10^5$ photon packages, which runs two minutes for a large galaxy with $M_\star\approx10^{11}\,\mathrm{M}_\odot$ and eight seconds for a smaller galaxy with $M_\star\approx10^{8}\,\mathrm{M}_\odot$ on a 16-core machine.

With this `SKIRT' method we can calculate realistic 3D UV fields for IllustrisTNG galaxies taking dust attenuation into account. The largest modelling uncertainty in this approach lies in the modelling of the star-forming regions, since they require estimates of multiple parameters that are not directly available from the output of the cosmological simulation (see also \citet{Kapoor2021}; \citealt{Trcka2022}). We also tested an alternative template library for star-forming regions (described in Kapoor et al. in prep.) which requires parameters that are more readily estimated from the TNG50 output. We found that using this alternative template library typically yields integrated \HI masses that agree with the MAPPINGS-III templates within one percent. Hence, we consider the MAPPINGS-III templates to be sufficiently accurate for this work.

For the UV modelling of TNG50-2/TNG100 galaxies, we note that these lower-resolution simulations can have very few star particles with our sample selection criteria (down to seven star particles for TNG100). For these barely resolved objects SKIRT cannot reproduce a meaningful UV field. Hence we use the Diemer UV field as replacement for the SKIRT UV field for galaxies with stellar masses below $10^8\,\mathrm{M}_\odot$ for TNG50-2 and TNG100.\\

\subsection{Atomic and molecular hydrogen fractions}\label{sec:HI fraction}

The $z=0$ snapshot data of IllustrisTNG contain the total and neutral hydrogen fraction in gas cells. Ideally, the neutral hydrogen fraction in the simulation is determined using on-the-fly ionizing radiative transfer calculations. For IllustrisTNG, the hydrogen ionization is approximated using the ionizing background (\citealt{Faucher2009}) and gas self-shielding with the fitting formula from \citet{Rahmati2013a}. Note that the neutral fractions for star-forming gas cells are for the effective gas temperature and must be recomputed. Hence, we use the intrinsic IllustrisTNG value for the neutral hydrogen fraction only for non star-forming cells. For star-forming cells, we follow \citet{Diemer2018}\footnote{See Eqns. A1-A6 in \citet{Stevens2019}.} and calculate the mass fraction of cold gas in the two-phase model of \citet{Springel2003}. This cold gas at a fixed $T=1000\,\mathrm{K}$ is assumed to be fully neutral while the hot phase is fully ionized.

The breakdown of neutral hydrogen gas into its atomic and molecular constituents requires the use of partitioning schemes. A variety of partitioning schemes exist and have been used to postprocess cosmological simulations (see \citealt{Diemer2018} for a detailed comparison between the different partitioning schemes and how that influences the \HI and H$_2$ properties of IllustrisTNG galaxies). Empirical partitioning schemes based on observations of local galaxies (\citealt{Blitz2006}; \citealt{Leroy2008}) find that the ratio between molecular and atomic surface densities scales with the hydrostatic midplane pressure as a power law. As shown by \citet{Diemer2018}, estimates of the midplane pressure are only physically reasonable when computed from a projection (in face-on orientation). The hydrogen partitioning then happens on this 2D grid on a pixel-by-pixel basis, hence 3D information is lost and it is only possible to create \HI maps in face-on orientation (it is not possible to rotate the galaxies back into a random orientation). Since we consider \HI maps in random orientation (as the WHISP data to which we compare the simulated \HI maps consist of randomly oriented \HI maps), we omit the empirical partitioning schemes completely from this study.

Analytical studies of \citet{Krumholz2009}\footnote{The partitioning scheme of \citet{Krumholz2009} is incorporated in the MUFASA and SIMBA simulations, splitting the hydrogen gas into its atomic and molecular phases at simulation runtime.} (KMT09), \citet{Krumholz2013} (K13), and \citet{Sternberg2014} (S14) consider idealized gas geometries with H$_2$ formation-dissociation balance to obtain the atomic and molecular hydrogen fractions. The main input parameters for these partitioning schemes are the surface density of neutral hydrogen, the dust-to-gas ratio, and (only for K13)\footnote{For the partitioning schemes of KMT09 and S14 the UV field strength drops out of the equations due to assumptions about the CNM-WNM pressure equilibrium.} the UV field strength at 1000\,\AA. Lastly, \citet{Gnedin2011} (GK11) and \citet{Gnedin2014} (GD14) use high-resolution (mass resolution up to $10^3\,M_\odot$) ISM simulations of isolated galaxies, coupled to a chemical network including H$_2$ formation on dust grains and photodissociation from Lyman-Werner photons, to partition the hydrogen based on the same three main parameters as the analytical partitioning scheme of K13. More specifically, the GK11 and GD14 approach consists of two steps: The first step consists of a cosmological simulation that is ran until $z=4$. This simulation extends over five virial radii of a system that would evolve to a typical $M\approx10^{12}\,\mathrm{M}_\odot$ system. In a second step, the UV radiation field at 1000\,\AA\ and the dust-to-gas ratio are fixed to a grid of different values and the simulation is continued for 600\,Myr. This setup allows for fresh gas infall into the galaxy, but cannot model effects from the larger environment such as mergers or filamentary accretion streams.

In this study we use the simulation-calibrated GD14 recipe as our default partitioning scheme as it is an update of the GK11 recipe (adding self-shielding of H$_2$ from line overlap) and has the UV field as an input parameter. Furthermore, it is more directly applicable to cosmological simulations as it is a hydrodynamical simulation that models the various ISM phases and outputs the \HI and H$_2$ fractions averaged over larger patches of the galaxy, while the analytical partitioning schemes consider simpler geometries such as a spherical cloud of gas. For our results we display the default partitioning scheme and treat the other UV-dependent partitioning schemes (GK11, K13) as uncertainties, shown as shaded areas. For some results we also consider the UV-independent KMT09 and S14 models which are plotted separately. The application of all of these partitioning schemes to IllustrisTNG is described in detail in Appendix C of \citet{Diemer2018}, except for KMT09 which we present in Appendix~\ref{sec:KMT}.

The analytical and simulation-calibrated partitioning schemes used here have up to three main input parameters; the surface density of neutral hydrogen ($\Sigma_\mathrm{H\,\textsc{i}+H_2}$), the dust-to-gas ratio relative to the average Milky Way value ($D_\mathrm{MW}$), and the UV field strength at 1000\,\AA\, relative to the average Milky Way value ($U_\mathrm{MW}$). Our estimates for $U_\mathrm{MW}$ are described in Section~\ref{sec:U_MW}. As in previous studies (e.g. \citealt{Lagos2015}), we assume that the dust-to-gas ratio scales with metallicity (as found observationally, e.g. \citealt{Remy2014}; \citealt{DeVis2017}; \citealt{DeVis2019}) such that $D_\mathrm{MW}=Z/Z_\odot$ with $Z$ the metal mass fraction and $Z_\odot=0.0127$ the solar metallicity used in IllustrisTNG (\citealt{Asplund2009}).

To convert the mass density of neutral hydrogen ($\rho_\mathrm{HI+H_2}$, calculated using the mass fraction of neutral hydrogen) into a surface density a certain length scale is needed. Two strategies to estimate surface densities from the simulation output have emerged: Volumetric modelling using the Jeans approximation and projected modelling using line-of-sight integration (\citealt{Diemer2018}). In the volumetric modelling approach, the surface density of neutral gas is obtained using the Jeans length $\lambda_\mathrm{J}$:

\begin{equation}\label{eq:Jeans}
    \Sigma_\mathrm{H\,\textsc{i}+H_2}=\rho_\mathrm{H\,\textsc{i}+H_2}\cdot\lambda_\mathrm{J}=\rho_\mathrm{H\,\textsc{i}+H_2}\cdot\sqrt{\frac{\gamma(\gamma-1)u}{G\rho}}.
\end{equation}
In theory, the value of $\gamma$ changes depending on the molecular fraction. Since the Jeans approximation by itself is already a crude approach we refrain from iterating the partitioning scheme to reach a self-consistent value of $\gamma$ as in \citet{Stevens2019}. In the projected modelling approach the surface density is estimated by rotating the galaxies into face-on position and projecting the gas densities onto a 2D grid. The hydrogen partitioning is then performed on this 2D grid, making it impossible to rotate the galaxy back into random orientation as the 3D information is lost. As we consider \HI maps in random orientation, we stick to the volumetric approach in this study and discuss projected modelling in the context of the different UV field estimates in Appendix~\ref{sec:Projected modelling}. We remark that the Jeans approximation to calculate surface densities breaks down in some regimes. However, the Jeans approximation gives the same H{\,\sc{i}}/H$_2$ results on average as the projected modelling approach (\citealt{Diemer2018}).

The main output of our H{\,\sc{i}}/H$_2$ modelling consists of $U_\mathrm{MW}$ parameters for the SKIRT, Diemer and Lagos method as well as 11 different molecular mass fractions $f_\mathrm{mol}$\footnote{We use $f_\mathrm{mol}$ to denote the mass ratio of molecular to neutral hydrogen.} (three different UV fields for GK11, GD14 and K13, plus the UV-independent schemes of KMT09 and S14). We store these values for all gas cells within the spherical aperture\footnote{Technically, only the SKIRT UV field calculation requires an aperture. To simplify the comparison between the different models we employ the same spherical aperture to store $U_\mathrm{MW}$ and $f_\mathrm{mol}$ for all our models. This aperture is defined as the radius that contains $99.9\,\%$ of the neutral hydrogen gas mass bound to the subhalo.}, for all 12'431 galaxies within our TNG50 base sample. We emphasize that our default H{\,\sc{i}}/H$_2$-model consists of the GD14 partitioning scheme with the SKIRT UV field.

\subsection{Observational data: The WHISP survey}\label{sec:WHISPsurvey}

\begin{table}
    \centering
    \begin{tabular}{cccc}
         Dataset & Beam\,[arcsec] & Pixel size\,[arcsec] & Noise\,[$\mathrm{M}_\odot\mathrm{pc}^{-2}$]  \\ \hline
         High-res. & 14 & 5 & 0.61\\
         Medium-res. & 30 & 10 & 0.18\\
         Low-res. & 60 & 20 & 0.061
         \end{tabular}
    \caption{Main characteristics of the observational WHISP data. The angular extent of the beam is listed as its FWHM. The noise corresponds to 1-$\sigma$ noise in the \HI surface density map. The beam FWHM and noise levels are average values for the entire WHISP sample, taken from \citet{Zwaan2005}.}
    \label{tab:WHISP}
\end{table}

We use observational data from the WHISP survey to compare to simulated \HI maps from IllustrisTNG. The WHISP survey targets the 21-cm line of some 300 local galaxies across the Hubble sequence (\citealt{Hulst2001}), selected by their 21-cm flux density and their stellar disc size. This interferometric survey at the Westerbork Synthesis Radio Telescope (WSRT) provides spatially resolved 21-cm line profiles such that \HI surface density maps and kinematics can be extracted. In this study we only focus on the \HI surface density maps.

In the WHISP survey, each object is observed in three different resolutions, with an average FWHM of the beam of 14, 30, and 60 arcsec, respectively. The sensitivity of the observations increased from a typical root-mean square error of 3 mJy/beam to 0.8 mJy/beam when the instrument was upgraded, leading to an average surface density sensitivity of $0.18\,\mathrm{M}_\odot\mathrm{pc}^{-2}$ for the medium-resolution maps (\citealt{Zwaan2005}). The observational characteristics that are relevant for this work are summarized in Table~\ref{tab:WHISP}.

The field-of-view (FOV) of the WHISP observations is typically 0.25 square degrees, centered on the object of interest. A substantial amount of maps contain other galaxies, galaxy interactions or mergers. While a study of these objects would be interesting on its own (see \citealt{Holwerda2011II}; \citealt{Holwerda2011-4}), we choose to avoid this additional complexity for the present study and analyze galaxies in IllustrisTNG one-by-one (i.e. processing each subhalo individually, see Section~\ref{sec:IllustrisTNG}). Hence, we want to exclude observations that feature ongoing interactions or have other galaxies in their FOV.

Recently, \citet{Naluminsa2021} compiled a catalogue of 228 galaxies observed within WHISP, containing only isolated objects (in their FOV, this does not refer to the environment in general) with good data quality. This sample is representative of the full WHISP sample in terms of galactic morphology. \citet{Naluminsa2021} also supplemented the WHISP data with infrared data from the WISE survey, enabling estimates of the stellar masses and star-formation rates. This is very advantageous for our purpose as we can select WHISP-like galaxies in IllustrisTNG based on their \HI mass, stellar mass, and star-formation rate. Hence, we use the WHISP sample defined in \citet{Naluminsa2021} for this study. Additionally, we remove two objects (UGC4483 and UGC9128) as their stellar masses are below our IllustrisTNG selection threshold of $10^7\,\mathrm{M}_\odot$, ending up with 226 galaxies in the observational galaxy sample.

We retrieve the moment-zero \HI maps for these 226 galaxies in all three available resolutions from the online Westerbork database\footnote{\url{http://wow.astron.nl/}}. The moment-zero maps, $\sum_\nu S_\nu$, are stored in mJy/beam ($S_\nu$ denotes the flux in a specific channel). We convert them to surface density maps assuming optically thin 21-cm emission (e.g. \citealt{Naluminsa2021}):

\begin{equation}\label{eq:WHISP}
\frac{\Sigma_\mathrm{H\,\textsc{i}}}{\mathrm{M_\odot pc^{-2}}}=8.85\frac{\sum_\nu S_\nu}{\mathrm{mJy/beam}}\frac{\Delta v}{\mathrm{km\,s^{-1}}}\Bigl(\frac{\Delta\alpha\cdot\Delta\delta}{\mathrm{arcsec}^2}\Bigr)^{-1},
\end{equation}
where $\Delta v$ is the velocity channel width and $\Delta\alpha$ ($\Delta\delta$) denote the FWHM of the major (minor) axes of the beam. We remark that for 47 WHISP galaxies the beam FWHM information was missing, for these galaxies we used the information from the online database for the highest-resolution observations and average values of $30\times30$ ($60\times60$) arcsec$^2$ for the medium (low) resolution observations, respectively.

To select WHISP-like galaxies from IllustrisTNG, we need some additional parameters to describe the WHISP galaxies like the \HI mass, stellar mass, and star-formation rate. Calculating these quantities requires a distance estimate. While \citet{Naluminsa2021} added distance estimates from the NASA Extragalactic Database (NED\footnote{\url{https://ned.ipac.caltech.edu/}}) to their galaxy catalogue, a substantial fraction is derived from redshifts, which can be unreliable for the local WHISP galaxies with a median distance of $\approx20\,\mathrm{Mpc}$ (\citealt{Zwaan2005}). Hence, we opt to compile our own distances from NED, using the most recent redshift-independent distance estimate for each galaxy. If such estimates are not available we use the distance estimated from the redshift, corrected for the influence of the Virgo cluster, the Great Attractor, and the Shapley Supercluster.

For each WHISP galaxy we compute the \HI mass from the distance estimate and the \HI surface density map (Eqn.~\ref{eq:WHISP}). We use the medium-resolution maps of the WHISP galaxies to compute their \HI masses, we note that the map resolution hardly affects the \HI masses. For the stellar masses and star-formation rates, we exactly follow the steps outlined in section 3 of \citet{Naluminsa2021} using their compilation of WISE fluxes but our own distance estimate compilation.

\subsection{Creation of HI maps}\label{sec:Maps}

\begin{figure*}
    \centering
    \includegraphics[width=\textwidth]{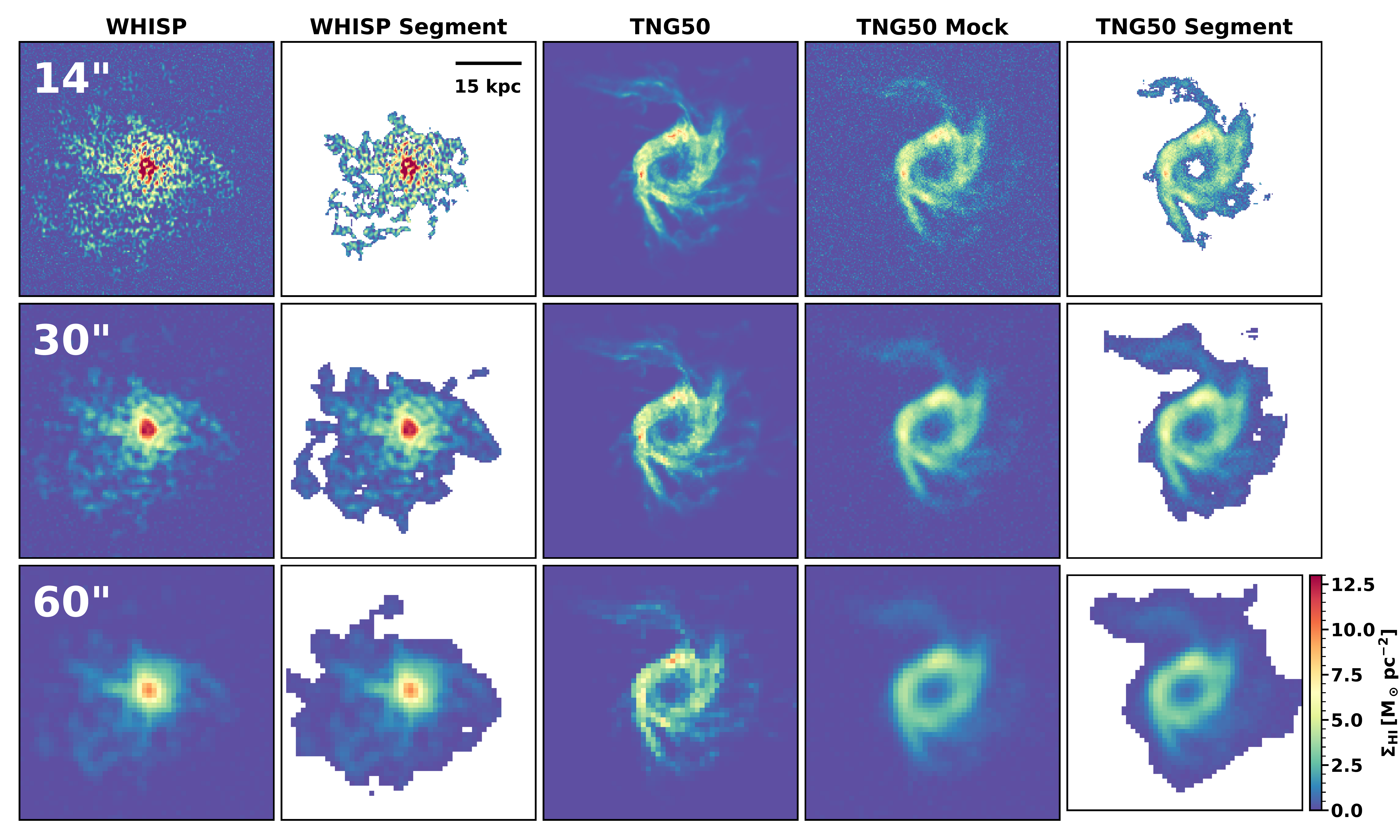}
    \caption{Visualization of various steps in our algorithms to create WHISP/mock IllustrisTNG \HI surface density maps. The different rows correspond to the three different map resolutions. The physical scale is the same in all images and indicated by the black 15\,kpc line. `WHISP': Example \HI map from WHISP (UGC528), where we replaced the NaN background pixels with Gaussian noise (after step (i) in Section~\ref{sec:WHISPmaps}). `WHISP Segment': The same WHISP \HI map, cutting out the object identified by the segmentation map (after step (ii) in Section~\ref{sec:WHISPmaps}). `TNG50': \HI map for the TNG50 galaxy that matches UGC528 best according to our algorithm (subhalo ID: 632310). To create the map we put the galaxy at the same distance and inclination as UGC528 (i.e. this map is generated after step (iii) in Section~\ref{sec:MockMaps}). `TNG50 Mock': The same TNG50 \HI map, after convolution with a WHISP-like beam and adding noise (after step (v) in Section~\ref{sec:MockMaps}). `TNG50 Segment': The same TNG50 \HI map, cutting out the object identified by the segmentation map (after step (vi) in Section~\ref{sec:MockMaps}).}
    \label{fig:ExampleImage}
\end{figure*}

\begin{figure*}
    \centering
    \includegraphics[width=\textwidth]{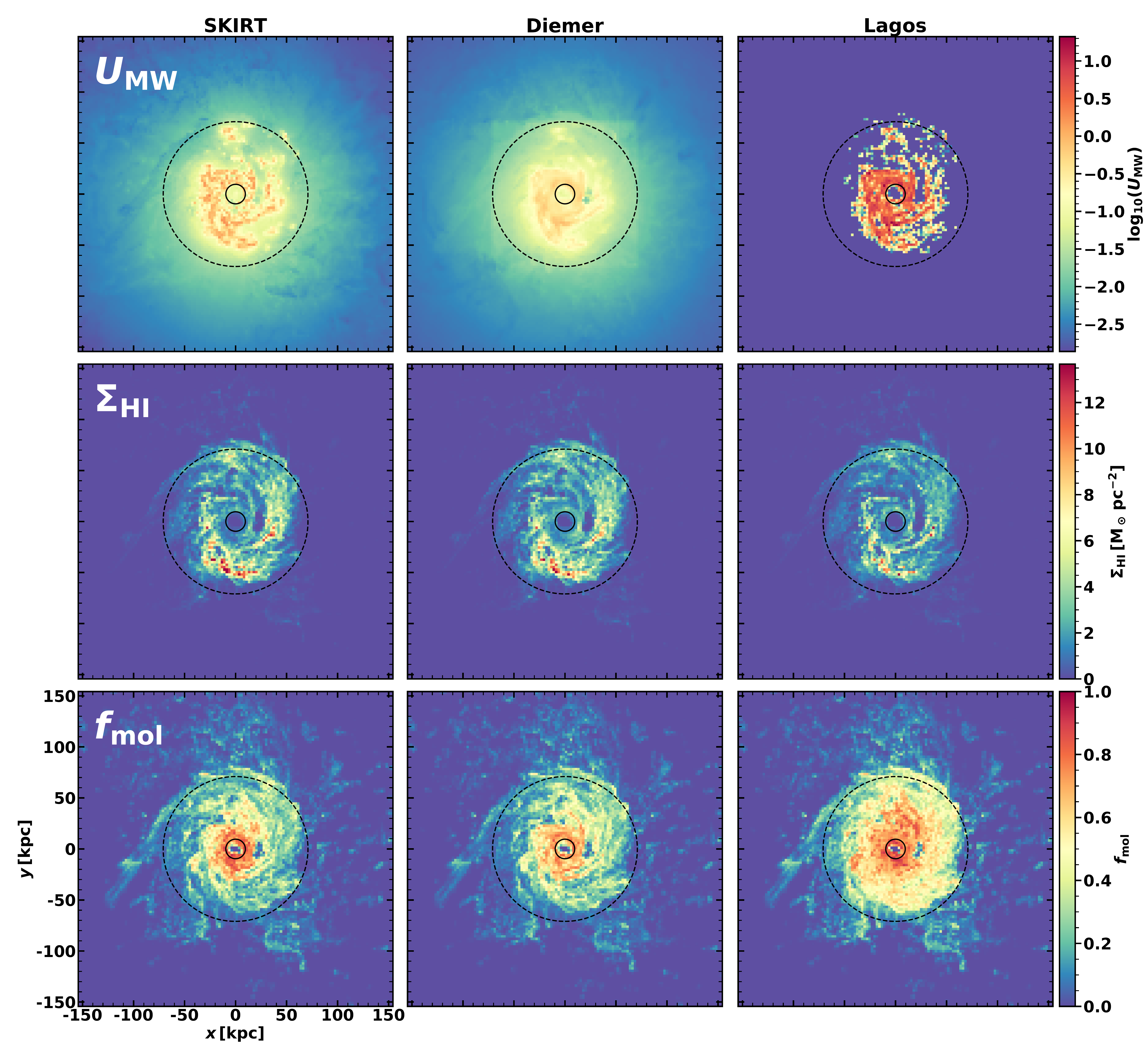}
    \caption{Maps of the UV fields (upper panels), H{\,\sc{i}} surface densities (middle panels), and molecular gas fractions (lower panels) for an example TNG50 galaxy at $z=0$ (subhalo ID: 474008). The atomic/molecular mass fractions are calculated with our default partitioning scheme (GD14). The different columns correspond to three different schemes to calculate the $U_\mathrm{MW}$ parameter. The black circles indicate the stellar (small, solid) and gas (large, dashed) half-mass radii (note that this includes all gas cells bound to the subhalo, not just \HI{}/H$_2$). We used the plain map algorithm (Section~\ref{sec:PlainMaps}) with random orientation to generate all maps. For the $f_\mathrm{mol}$ maps we weighted each gas cell by its neutral hydrogen mass.}
    \label{fig:ExampleMap1}
\end{figure*}

Here we describe how we generate \HI maps for the simulation and observational galaxy samples. Part of the analysis of this paper compares different simulation postprocessing methods against each other. For these parts, we use a fairly basic algorithm to create `plain' \HI maps for IllustrisTNG galaxies, described in Section~\ref{sec:PlainMaps}. When we compare IllustrisTNG against observational data, we use a more advanced algorithm to generate `mock' \HI maps (described in Section~\ref{sec:MockMaps}) with the aim of emulating the most important steps of the observations. Lastly, we also make some minor adjustments to the observational WHISP \HI maps, described in Section~\ref{sec:WHISPmaps}. The main steps of the algorithm to create WHISP and mock IllustrisTNG \HI maps are visualized in Figure~\ref{fig:ExampleImage}.

\subsubsection{IllustrisTNG: Plain maps}\label{sec:PlainMaps}

(i) Orientation: When creating plain maps in random orientation, we leave the simulated galaxy in its intrinsic orientation. For plain maps in face-on orientation, we rotate the galaxies such that its angular momentum vector aligns with the $z$-axis of the simulation box. Following \citet{Diemer2018} we calculate the galactic angular momentum using all gas cells within the 3D gas half-mass radius. If there are fewer than 50 gas cells within this radius we use all star particles within two stellar half-mass radii instead.\\

(ii) Projection: We always project the gas cells along the $z$-axis of the simulation box. We employ a consistent projection algorithm for these tasks, specifically we use the method that was introduced by \citet{Diemer2018} which we briefly summarize here. Since the moving-mesh cells of IllustrisTNG have complex shapes we apply an adaptive Gaussian smoothing kernel to the gas cells. For the smoothing kernel, we use a width of $\sigma=0.5(m/\rho)^{1/3}$. For the calculation of the \HI surface density, the algorithm simply sums up the \HI masses of all smoothed gas cells in each pixel and divides by the pixel area. The map size is set to lie within the spherical aperture such that the map side length is $\sqrt{2}\cdot r_\mathrm{A}$. We resolve the plain maps by a fixed number of pixels ($128\times128$). 

In some cases we are also interested in maps of other quantities than \HI surface density. We employ the same projection algorithm in these cases. For quantities that are averaged (instead of summed) within each pixel (e.g. $U_\mathrm{MW}$, $D_\mathrm{MW}$) we weigh each gas cell by its total mass unless otherwise noted.

\subsubsection{IllustrisTNG: Mock maps}\label{sec:MockMaps}

A fair comparison between observations and cosmological simulations needs the two galaxy samples to be broadly comparable. Furthermore, it requires some sort of mock-observation routine which mimics the observational procedure. We do not attempt to fully emulate the interferometric observations of WHISP but include the three most important effects: Mimicking the angular pixel sizes of the observations, smoothing with the beam of the instrument, and adding noise. This is a similar strategy as is applied in the MARTINI tool to create mock \HI datacubes from cosmological simulations (\citealt{Oman2019}), although there is no noise incorporated in MARTINI.\\

(i) Selection: For each WHISP galaxy, we select IllustrisTNG galaxies with similar stellar masses ($\pm0.2\,\mathrm{dex}$) and star-formation rates ($\pm0.3\,\mathrm{dex}$). The tolerances for these quantities correspond to the 2-$\sigma$ error bars adopted by \citet{Naluminsa2021}. For $\approx20\,\%$ of the WHISP galaxies in the \citealt{Naluminsa2021} sample, no SFR estimate is available because they were not detected in the WISE W3 band. For these galaxies we only used the stellar mass criterion. In a second step, we then select the five IllustrisTNG galaxies that are closest to the WHISP galaxy in terms of $\log_{10}M_\mathrm{HI}$ out of the galaxies that are similar in terms of stellar mass and star-formation rate. The factor of five is chosen to increase the statistics of the simulated galaxy sample while keeping the \HI mass distribution similar. As we find at least five similar IllustrisTNG galaxies (in terms of stellar mass and SFR) for each WHISP galaxy (typically dozens of IllustrisTNG galaxies conform to the stellar mass-SFR selection criteria), the `mock' TNG50 sample consists of $5\cdot226=1130$ galaxies. We remark that $\approx20\,\%$ of this sample is just selected by $M_\star$ and $M_\mathrm{H\,\textsc{i}}$ due to some WHISP galaxies being undetected in the WISE W3 band. We have verified that these TNG50 galaxies typically have low star-formation rates, as expected from their observational counterparts for which we only have an upper bound on the SFR.\\

(ii) Orientation: We match the inclinations of the IllustrisTNG galaxies to their corresponding WHISP galaxies, by rotating IllustrisTNG galaxies such that the angle between the $z$-axis of the simulation box and the angular momentum vector matches the WHISP galaxy inclination. We retrieve the inclinations of the WHISP galaxies from the galaxy catalogue of \citet{Naluminsa2021}.\\

(iii) Projection: We put the IllustrisTNG galaxies at the distances of their corresponding WHISP galaxies. The galaxies are then projected along the $z$-axis of the simulation onto a 2D grid. Each map is created in three different resolutions as in the observational data, the angular pixel sizes (5, 10, and 20 arcsec, respectively) are chosen to match the WHISP data. We use the same map size as in Section~\ref{sec:PlainMaps} of $\sqrt{2}\cdot r_\mathrm{A}$.\\

(iv) Smoothing: To emulate the effect of the WHISP beam we then convolve these \HI maps with a circular 2D Gaussian kernel, using the average WHISP beam FWHM (14, 30, and 60 arcesonds, respectively) for the three different resolutions.\\

(v) Noise: After beam convolution, we add Gaussian noise to the \HI maps. We use the average 1-$\sigma$ WHISP noise levels (see Table~\ref{tab:WHISP}), which vary depending on the map resolution.\\

(vi) Segmentation: For the calculation of the morphological parameters and the CDDF, the main object needs to be identified and separated from the background in the \HI map. While we always project only one subhalo per IllustrisTNG \HI map, it is possible that a subhalo consists of a neutral gas disk and additional disconnected neutral gas structures. For our analysis, we only want to consider the main object (in this case the neutral gas disk), and hence need to segment the \HI map. We use the \texttt{photutils}\footnote{We use the \texttt{photutils.detect\_sources} function with the average 1-$\sigma$ noise (see Table~\ref{tab:WHISP}) as threshold and five pixels as the minimum number of connected pixels.} \texttt{python} package to find the object which belongs to the central pixel. If the central pixel is assigned to the background we instead use the largest object in the \HI map. To make the segmentation map which cuts out the galaxy of interest more regular we smooth it with a uniform boxcar filter measuring five pixels in each dimensions. By visual inspection, we find that this algorithm reliably identifies the galaxy in the \HI map.

\subsubsection{WHISP: Observational maps}\label{sec:WHISPmaps}

The extraction of \HI maps for the WHISP galaxies is described in Section~\ref{sec:WHISPsurvey}. We perform two additional steps to generate the final WHISP \HI maps that we analyze in this study.

(i) Noise: The WHISP \HI maps are inherently 3-$\sigma$-clipped, meaning that they do not contain any background (all background pixels are set to NaN values). This is problematic for the calculation of the morphological parameters, as some of the parameters need a background in the image in order to be accurately computed. Hence, we replace all NaN pixels from the WHISP \HI maps with Gaussian noise, using the average 1-$\sigma$ WHISP noise levels depending on map resolution.\\

(ii) Segmentation: We create segmentation maps for the WHISP galaxies using exactly the same algorithm as described in Section~\ref{sec:MockMaps}. Again, we find by visual inspection that this algorithm reliably identifies the main galaxy in the FOV of the observation.

\section{UV field comparison}\label{sec:UV comparison}

We start our analysis by assessing the differences in the SKIRT, Diemer, and Lagos UV field estimates, and explore how this affects the \HI and H$_2$ properties of the simulated galaxy population in TNG50.

To qualitatively compare the different UV field estimates we show 2D projections of the $U_\mathrm{MW}$ parameter (calculated using the plain map algorithm in random orientation, see Section~\ref{sec:PlainMaps}) for an example TNG50 galaxy in the top panels of Figure~\ref{fig:ExampleMap1}. This galaxy has a stellar mass of $9.3\times10^{10}\,\mathrm{M_\odot}$, a high gas-to-stellar mass ratio of 2.1, and a star-formation rate of $9.0\,\mathrm{M_\odot yr^{-1}}$. The 3D stellar and gas half-mass radii are shown as solid and dashed circles, respectively. The angular momentum vector of the galaxy is almost parallel to the $z$-axis of the simulation, hence the random projection in Figure~\ref{fig:ExampleMap1} displays an almost face-on orientation of the galaxy.

For the SKIRT UV field (top left panel), some clumpy structure in the spiral arms due to the star-forming regions (young star particles with ages less than 10\,Myr) that strongly emit UV is visible. In the outskirts (outside the gas half-mass radius), the attenuation and scattering by dust grains imprints some structure on the UV field. We remark that in the SKIRT UV field, radiation from the evolved stellar population is also modelled, unlike in the other UV field models. For the Diemer UV field (top center panel), radiation is emitted from all star-forming gas cells. These cells are more common but less luminous than the star-forming regions in the SKIRT recipe (star particles with ages below 30\,Myr), leading to a less clumpy UV field in the spiral arms for the Diemer UV field. Since the radiation is transported without considering absorption/scattering the radiation field is very smooth in the galactic outskirts as well. We remark that there is a minor discontinuity in the Diemer UV field at the edges of a square enclosing the gas half-mass radius. This is due to numerical resolution issues in the calculation of the Diemer UV field as the algorithm splits the galaxy into a high-resolution region within the gas half-mass radius and a low-resolution region outside of it, allowing a significant speedup of the algorithm. We do not expect that this affects any of our results. Lastly, the Lagos UV field (top right panel) scales the UV flux to the star-formation rate of gas cells. Since there is no transport of radiation in this approach there is a strong bimodality between star-forming and quiescent gas cells. The gas cells without star formation only receive the homogeneous background UV field of $U_\mathrm{MW}=0.00137$ (\citealt{Faucher2009}), note that this floor is applied to all three UV field estimates. 

We also show maps of \HI{} surface densities and molecular fractions calculated with the three different UV field estimates in Figure~\ref{fig:ExampleMap1} (using our default partitioning scheme of GD14). Despite significant differences in the UV field, the $\Sigma_\mathrm{H\,\textsc{i}}$ and $f_\mathrm{mol}$ maps for this example galaxy appear almost identical for the SKIRT and Diemer UV field estimates. We remark that the neutral gas hole in the center of this galaxy is due to feedback from the supermassive black hole (see Section~\ref{sec:Holes}).

In Section~\ref{sec:Mass Functions} and Section~\ref{sec:Profiles} we examine whether the different UV field estimates lead to significant statistical discrepancies for ensemble quantities of the entire galaxy population (within our base sample of 12'431 galaxies) in TNG50.

\subsection{Radial profiles}\label{sec:Profiles}

\begin{figure}
    \centering
    \includegraphics[width=\columnwidth]{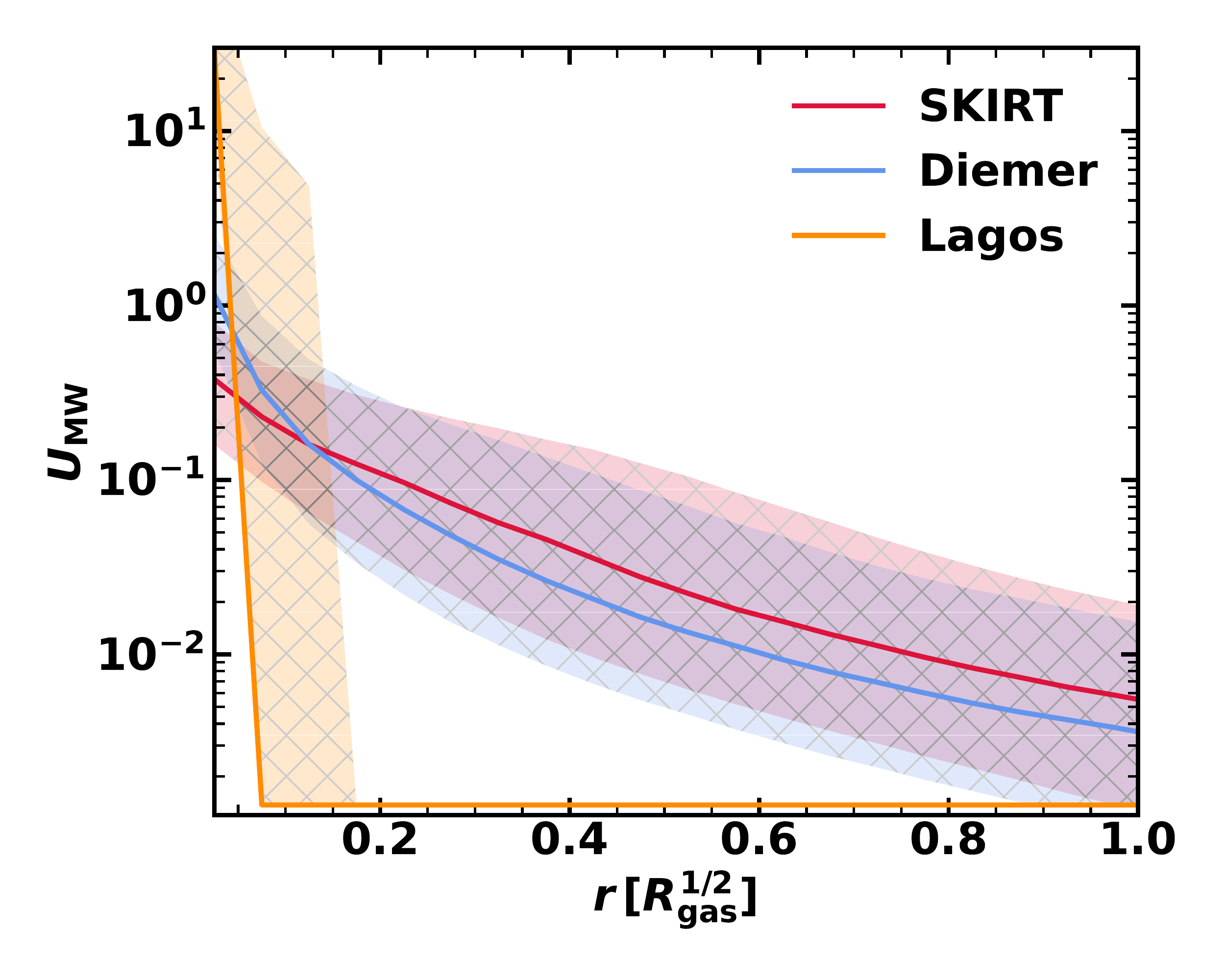}
    \caption{Median radial profiles for the UV field parameter $U_\mathrm{MW}$, considering only galaxies with $M_\mathrm{H_2}>10^8\,\mathrm{M}_\odot$. The galaxies are projected onto a face-on orientation and stacked by their gas half-mass radius. The lines indicate the median UV profiles, hatched areas show the interquartile range.}
    \label{fig:UVprofile}
\end{figure}

\begin{figure}
    \centering
    \includegraphics[width=\columnwidth]{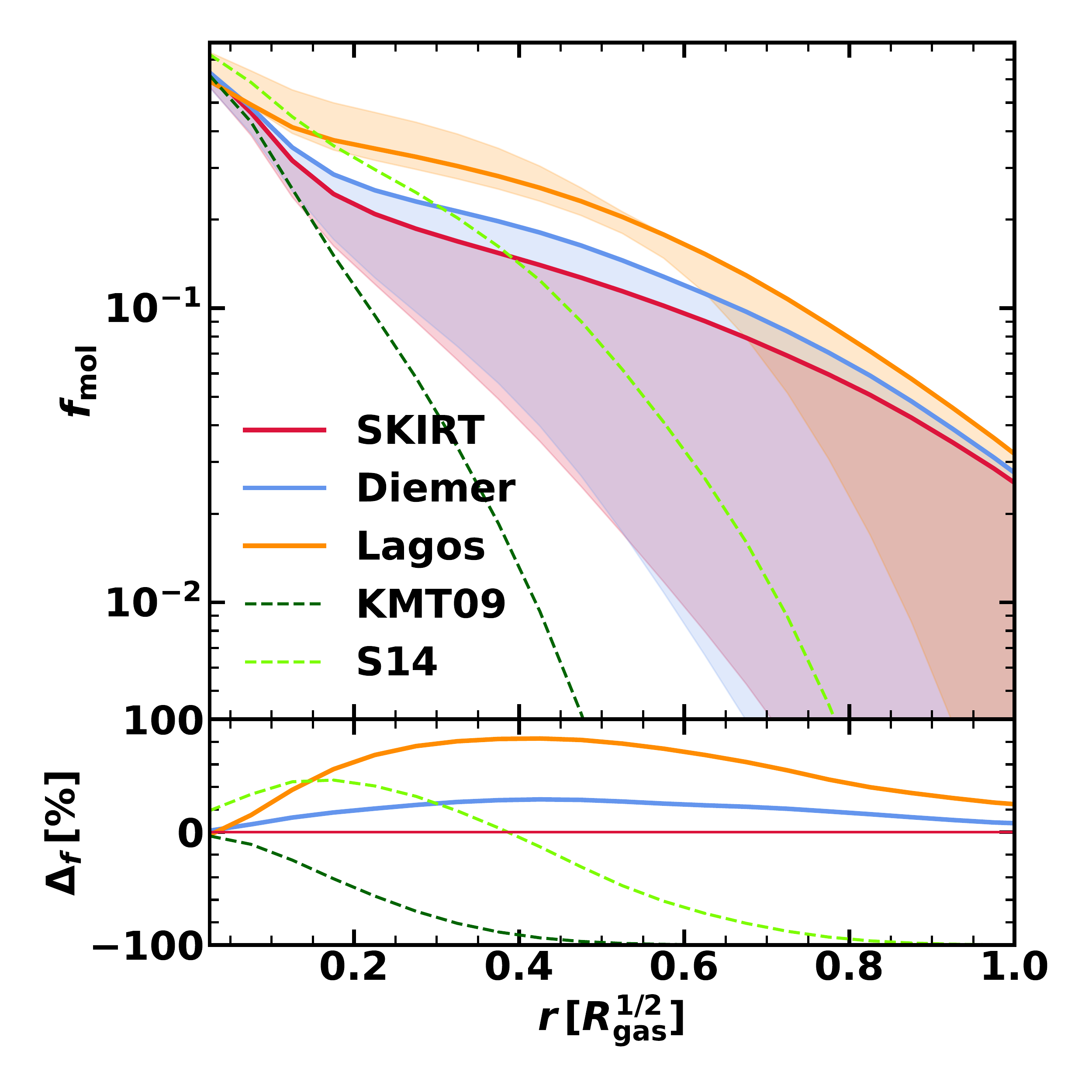}
    \caption{Median radial profiles for the molecular hydrogen mass fraction $f_\mathrm{mol}$. We only consider galaxies with $M_\mathrm{H_2}>10^8\,\mathrm{M_\odot}$ here. The galaxies are projected onto a face-on orientation and stacked by their gas half-mass radius. Solid lines indicate the default partitioning scheme (GD14), the spreads between the UV-dependent partitioning schemes are visualized as shaded areas. The UV-independent models are shown as dashed lines (KMT09 and S14). The bottom panel shows the difference in the $f_\mathrm{mol}$ profiles relative to the default model. For a clearer visualization we smoothed all curves with a Gaussian filter with $\sigma=0.05\,R_\mathrm{gas}^{1/2}$.}
    \label{fig:Profile}
\end{figure}

To assess the impact of the UV field on the H{\,\sc{i}}-H$_2$-transition on the galaxy population in a more statistical fashion, we calculate 1D projected radial profiles of UV fields and molecular gas fractions for the TNG50 base sample in this section. Following \citet{Diemer2018}, we only consider galaxies with $M_\mathrm{H_2}>10^8\,\mathrm{M}_\odot$ (we use our default \HI model here to calculate the mass for the galaxy selection) in this section as the $f_\mathrm{mol}$ profiles of galaxies with less molecular gas are noisy. The exact value of this cutoff does not affect our conclusions.

We show median $U_\mathrm{MW}$ radial profiles in Figure~\ref{fig:UVprofile}. The interquartile ranges are shown as hatched areas. For the calculation of the radial profiles we first rotated each galaxy into a face-on orientation (see step (i) of Section~\ref{sec:PlainMaps}) and considered 2D radial profiles (i.e. using the projected distance to the galaxy center, not the 3D distance). For each radial bin, we then compute the median of the $U_\mathrm{MW}$ parameter of all gas cells within this bin (using the mean instead of the median would significantly bias $U_\mathrm{MW}$ for the Lagos UV field, as star-forming gas cells receive UV fluxes that are 4-6 orders of magnitude larger than non star-forming cells). If there are no gas cells for a galaxy within a radial bin $U_\mathrm{MW}$ is set to the uniform UV background. To compute a single average profile for the entire galaxy sample we stacked the galaxies by their gas half-mass radii and computed the median profile.

At the center the Lagos UV field exceeds the other estimates by more than an order of magnitude due to the large unattenuated UV field in star-forming gas. As the Lagos UV field is scaled to the gas cell star-formation rate and most of the gas cells at larger radii do not form stars, the Lagos UV field quickly decreases to the uniform UV background at $U_\mathrm{MW}=0.00137$. The Diemer UV field has a steeper UV profile than the SKIRT field, which is due to the centrally concentrated star-forming gas cells being the only UV sources in the Diemer UV field. The SKIRT UV field is more extended as we treat all star particles as UV sources (see Table~\ref{tab:UVfields}).

\begin{figure*}
    \centering
    \includegraphics[width=\textwidth]{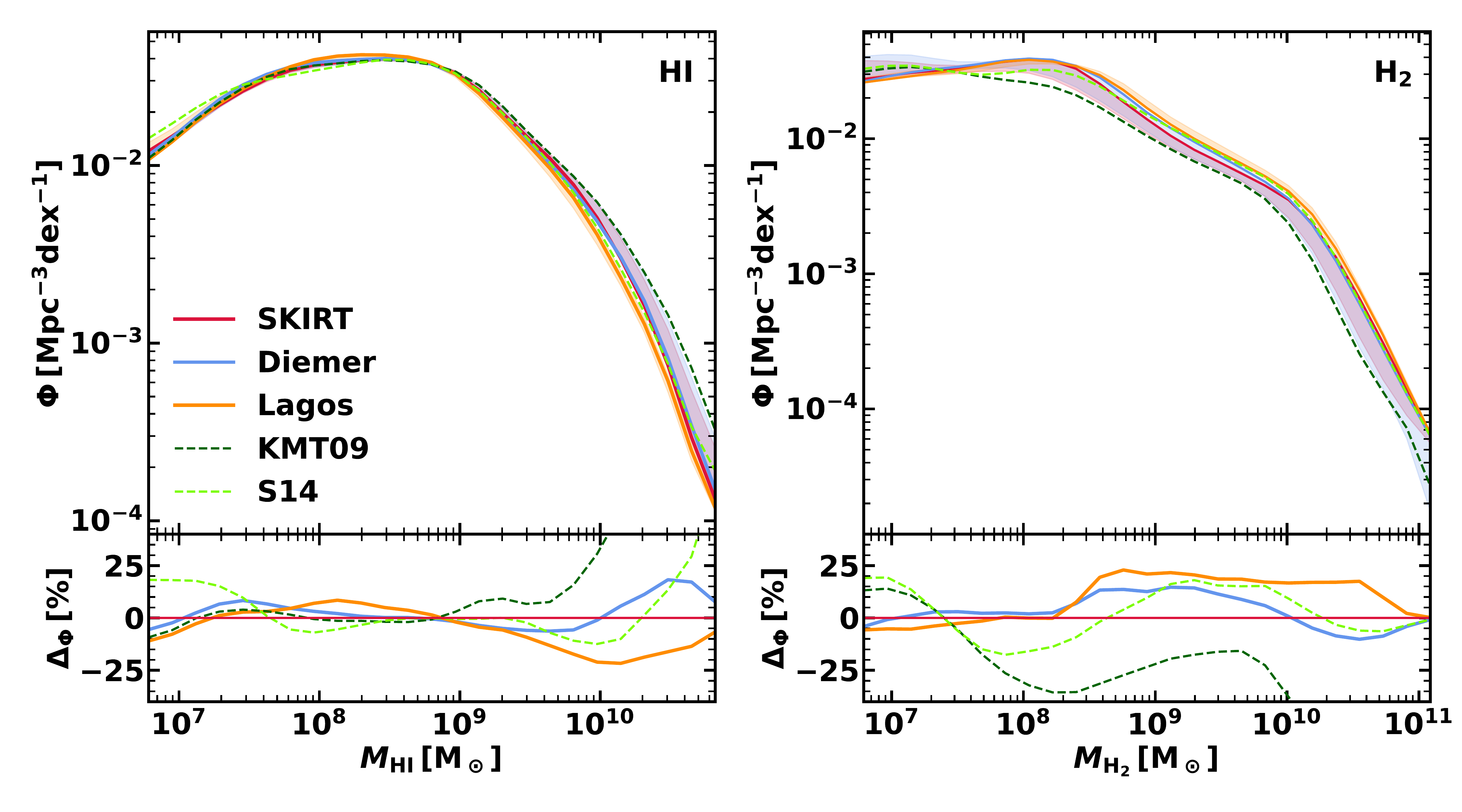}
    \caption{\HI (left) and H$_2$ (right) mass functions for our TNG50 galaxy sample. We show the different partitioning schemes as in Figure~\ref{fig:Profile}: Solid lines show the mass functions for the three UV models with the default scheme. The shaded areas indicate the spread due to the different UV-dependent partitioning schemes. The two UV-independent schemes are shown as dashed lines (KMT09 and S14). The differences between the mass functions relative to the default H{\,\sc{i}}/H$_2$ (GD14 with the SKIRT UV field) model are shown in the bottom panels. For a clearer visualization we smoothed all curves with a Gaussian filter with $\sigma\approx0.17\,\mathrm{dex}$.}
    \label{fig:MassFunction}
\end{figure*}

We compare the different calculations for the H$_2$ mass fractions by plotting the median projected radial profiles of $f_\mathrm{mol}$ in Figure~\ref{fig:Profile}. For each galaxy, we calculate its profile by computing the mean molecular fraction in each radial bin, similar to the calculation of the UV field profiles. If there are no gas cells for a galaxy within a radial bin, $f_\mathrm{mol}$ is set to zero. We find that if we instead ignore these radial bins for the calculation of the median profiles $f_\mathrm{mol}$ increases nonphysically for large radii. As for the UV field profiles, we stack the profiles of all galaxies by their half-mass radius and compute the median $f_\mathrm{mol}$ profile of the galaxy population. Applying the mean instead of the median to the stacked profiles (as is done in \citealt{Diemer2018}) gives higher $f_\mathrm{mol}$ at larger radii and reduces the scatter between the different models substantially. As the mean profiles are dominated by few galaxies that have high molecular fractions at larger radii and for consistency with Figure~\ref{fig:UVprofile} we opt for median profiles in Figure~\ref{fig:Profile}. For the three UV models (SKIRT, Diemer, Lagos) we show the default hydrogen partitioning scheme (GD14) as solid lines and the spread between the UV-dependent schemes as shaded areas. We also show the UV-independent partitioning schemes (KMT09 and S14). The differences in the molecular fractions relative to the default \HI model (GD14 with the SKIRT UV field) are shown in the bottom panels.

The significantly lower Lagos UV field (except in the center) manifests in higher molecular fractions. For the SKIRT and Diemer $f_\mathrm{mol}$ profiles, the spread due to the different partitioning schemes is much larger than the difference due to changing the UV field. The discrepancy in the $f_\mathrm{mol}$ profiles comparing the different UV fields reaches its maximum at $\approx0.4\,R_\mathrm{gas}^{1/2}$. This is not particularly surprising because $f_\mathrm{mol}$ is 20-40\,\% at this radius, so this is broadly the region at which the H{\,\sc{i}}-H$_2$-transition takes place. The H{\,\sc{i}}-H$_2$-transition occurs typically over a very narrow range in densities, and other parameters like $U_\mathrm{MW}$ can slightly shift this density range (see e.g. Figure 1 of \citealt{Gnedin2011}). This means that the gas density in this region is in a range where the partitioning schemes are particularly susceptible to changes in the UV field. Lastly, we remark that the default GD14 partitioning scheme minimizes the differences in the molecular fractions upon variation of the UV field. Hence, we expect more significant discrepancies in the H{\,\sc{i}}/H$_2$ properties of galaxies between different UV field estimates when using other partitioning schemes (GK11 or K13).

\subsection{HI and H2 mass functions}\label{sec:Mass Functions}

We consider H{\,\sc{i}}/H$_2$ mass functions for our base sample in the upper panels of Figure~\ref{fig:MassFunction}. As in Figure~\ref{fig:Profile} we display the UV-dependent partitioning schemes by showing the default partitioning scheme and spread between schemes for each UV field estimate individually (solid lines and shaded areas). Differences in the molecular fraction relative to the default model are shown in the bottom panel.

The \HI mass function (HIMF) is very robust against variations of the partitioning schemes and UV field estimates, except for the high-mass end. \HI masses calculated using the Lagos UV field are lower compared to the SKIRT and Diemer UV fields, with discrepancies in the HIMF up to 20\,\%. For the H$_2$ mass function (H2MF), discrepancies are similar but begin at much lower masses. The differences between the partitioning schemes are generally larger for H$_2$ than for \HI. For both mass functions, the default GD14 partitioning scheme is the \HI model with the least tension between the different UV models (as seen in Figure~\ref{fig:Profile}). Still, the shaded areas between the different UV models do not overlap for high \HI or intermediate H$_2$ masses, indicating that the discrepancy between different UV models is robust to the choice of the partitioning scheme. For the UV-independent partitioning schemes, we remark that the S14 model is in good agreement while the KMT09 model significantly exceeds other HIMF model predctions at the high-mass end (and vice versa for the H2MF).

We note that our \HI mass functions are not directly comparable to the results of \citet{Diemer2019} due to different sample definitions: \citet{Diemer2019} consider all galaxies above a certain gas \textit{or} stellar mass threshold, while we need a minimal stellar mass for all galaxies for SKIRT to work properly. This means we miss galaxies with low stellar mass ($M_\star<10^7\,\mathrm{M_\odot}$) and high gas fractions when calculating the gas mass functions. Since the H{\,\sc{i}}-to-stellar mass fraction in TNG50 at a stellar mass of $\sim10^7\,\mathrm{M_\odot}$ reaches values up to $\sim10$ (\citealt{Diemer2019}, figures including the TNG50 simulation can be found online\footnote{\url{benediktdiemer.com/data}}), the calculated \HI mass function is complete only above an \HI mass of $\sim10^8\,\mathrm{M_\odot}$. Indeed, we find that our HIMF agrees with the one from \citet{Diemer2019} above \HI masses of $2\times10^8\,\mathrm{M_\odot}$. For molecular hydrogen the stellar mass cutoff is irrelevant: Because the H$_2$-to-stellar mass fractions are always smaller than unity, a stellar mass cutoff at $M_\star=10^7\,\mathrm{M_\odot}$ does not exclude any galaxies with $M_\mathrm{H_2}>10^7\,\mathrm{M_\odot}$. Hence, our $\mathrm{H}_2$ mass function calculated with the Diemer UV field matches the result from \citet{Diemer2019}.

\begin{figure*}
    \centering
    \includegraphics[width=\textwidth]{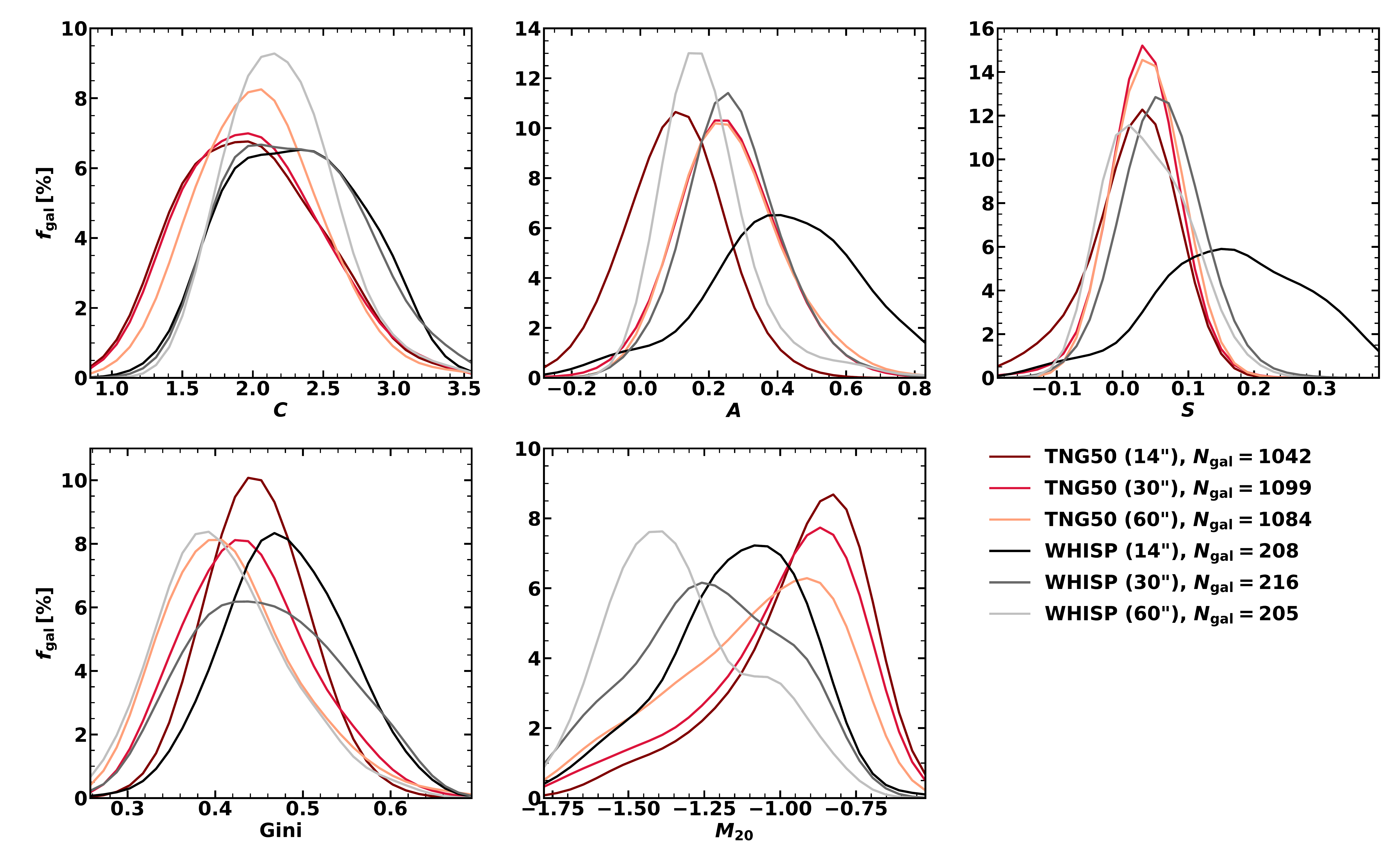}
    \caption{Histograms of non-parametric morphologies (concentration, asymmetry, smoothness, Gini, and $M_{20}$ statistics) for the WHISP and TNG50 \HI maps, at varying angular resolution. We only use the default \HI model for the TNG50 \HI maps here. The sizes of the galaxy samples vary slightly with resolution as \texttt{statmorph} does not always run successfully, depending on the resolution. For a clearer visualization we smoothed the histograms with a Gaussian filter with $\sigma$ equal to $6.7\,\%$ of the $x$-axis stretch.}
    \label{fig:MorphologiesVarResWISE} 
\end{figure*}

\section{Non-parametric morphologies of HI maps}\label{sec:morphologies}

To test our \HI model and to evaluate the realism of the spatial distribution of \HI gas within TNG50 galaxies, we compare simulated \HI surface density maps to 21-cm maps from the observational WHISP dataset. We use the full mock map algorithm (Section~\ref{sec:MockMaps}) to generate a WHISP-like sample of 1130 TNG50 \HI maps, to be compared to 226 WHISP \HI maps. To quantitatively compare the \HI maps of the two samples we consider non-parametric morphologies (concentration, asymmetry, smoothness, Gini and $M_{20}$ statistics). We calculate these statistics for the TNG50 and WHISP maps consistently with the \texttt{statmorph} tool (\citealt{Rodriguez-Gomez2019}). 

\subsection{Statmorph}

\texttt{statmorph} is a \texttt{python} tool for parametric and non-parametric analyses of astronomical images, and has already been used to study mock images in optical bands for IllustrisTNG (\citealt{Rodriguez-Gomez2019}; \citealt{Guzman2022}). Besides the image, the \HI surface density map in our case, \texttt{statmorph} requires a segmentation map and an estimate of the noise as input. A segmentation map separates all objects in the image from the background. As we process galaxies individually in IllustrisTNG and use a galaxy sample from WHISP that only contains isolated objects, we have only one object per image to analyze. The segmentation map is calculated as described in Section~\ref{sec:Maps}.

\texttt{statmorph} also requires an estimate of the noise on the maps. We use the simpler gain option of \texttt{statmorph} which calculates Poisson noise, the non-parametric morphologies are independent of the actual gain value. For $\approx\!5\%$ of the galaxies\footnote{This fraction is very similar for the WHISP and the TNG50 galaxy samples individually.}, \texttt{statmorph} does not finish properly, mostly due to irregular segmentation maps. Such galaxies are omitted from our analysis. We find that the results at medium resolution are most reliable, as they pick up more features than the low-resolution maps and \texttt{statmorph} successfully runs more often than for the high resolution. Hence, we consider the medium-resolution \HI maps of 30\,arcsec as default in the following analysis.

\subsection{Morphological results}

\begin{figure}
    \centering
    \includegraphics[width=1.05\columnwidth]{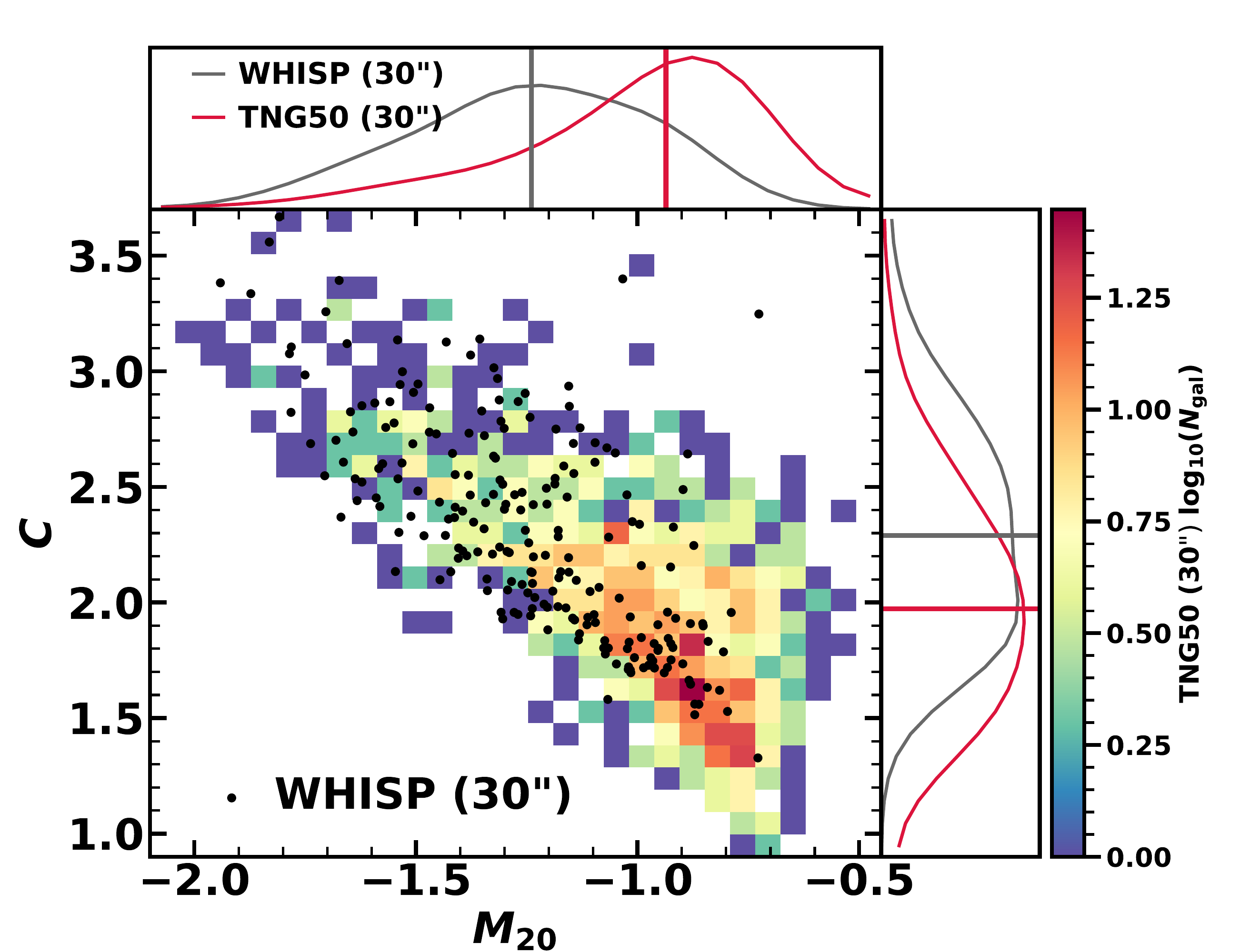}
    \caption{$M_{20}$-concentration relation for the WHISP (black data points) and TNG50 samples (colored 2D histogram) for the medium-resolution \HI maps. We only use the default \HI model for the TNG50 \HI maps here. The 1D histograms for the $M_{20}$ and $C$ statistics for the same galaxy samples are shown at the edges of the plot (smoothed as in Figure~\ref{fig:MorphologiesVarResWISE}), straight lines indicate the median values.}
    \label{fig:MorphologyScatter}
\end{figure}

We show histograms of the non-parametric morphologies for the TNG50 and WHISP \HI maps in varying angular resolution in Figure~\ref{fig:MorphologiesVarResWISE}. To avoid overcrowded figures we only show our default partitioning scheme (GD14 plus SKIRT UV field) in this section, using other partitioning schemes or UV fields yields comparable results. We remark that the number of galaxies plotted in the histograms has a small dependency on the angular resolution, since \texttt{statmorph} is not always able to measure the morphological parameters. We have verified that this does not affect the distributions in Figure~\ref{fig:MorphologiesVarResWISE}, i.e. the galaxies that are lost in the high-resolution TNG50 maps do not differ significantly in terms of the morphological statistics from the overall galaxy sample.

Comparing the WHISP and TNG50 medium-resolution (30\,arcsec) maps, we find that the asymmetry statistic matches well, the TNG50 smoothness statistic is slightly lower (i.e. the TNG50 \HI maps are slightly smoother than the WHISP maps), and the distribution of the Gini statistic is a bit more skewed in the WHISP sample. For the TNG50 maps, the scaling with angular resolution seems to be weaker for the smoothness statistic and reversed for the asymmetry statistic. As the maps at resolutions other than 30" are less reliable we do not investigate this discrepancy further. More significant differences arise in the concentration and $M_{20}$ statistics. Denoting the difference of the TNG50 and WHISP medians of a morphological statistic $k$ as $\Delta\tilde{k}$, we have more concentrated WHISP \HI maps ($\Delta\tilde{C}\approx-0.32$) and find $\Delta\tilde{M}_{20}\approx0.30$.

We further explore the deviations in the concentration and $M_{20}$ statistics in Figure~\ref{fig:MorphologyScatter} by plotting their correlation for the medium-resolution maps. This is also the strongest cross-correlation between non-parametric morphologies found by \citet{Holwerda2011II} and can be used to identify interacting galaxies as outliers in the top right corner. Due to the definitions of $C$ (increasing with brighter pixels in the galaxy center) and $M_{20}$ (increasing with brighter pixels in the outskirts), these statistics are usually anticorrelated. Such an anticorrelation is found in the WHISP data (see Figure~\ref{fig:MorphologyScatter} or \citealt{Holwerda2011-4}), in synthetic optical images of IllustrisTNG galaxies (\citealt{Rodriguez-Gomez2019}; \citealt{Guzman2022}), and in images across the UV-submm wavelength range for both observed and simulated galaxies (\citealt{Baes2020}; \citealt{Kapoor2021}; \citealt{Camps2022}).

We find that the two samples broadly follow the same anticorrelation, with the TNG50 galaxies shifted to the bottom right corner. Visual inspection of the \HI maps with low concentration and high $M_{20}$ reveals that those are exclusively face-on galaxies, while the TNG50 outliers at intermediate concentration and high $M_{20}$ are mostly edge-on. These angular momentum trends are less pronounced for the WHISP galaxies. We do not find any other obvious characteristic such as \HI mass that distinguishes the TNG50 outliers from the WHISP galaxies.

\subsection{Central HI holes in TNG50}\label{sec:Holes}

By visual inspection of the TNG50 and WHISP \HI maps, we find that many TNG50 galaxies feature large central \HI holes, while such \HI holes are much less prevalent in the WHISP data\footnote{On the other hand, by visual inspection of the 34 \HI maps from the THINGS \HI survey (\citealt{Walter2008}) it seems that central \HI holes are prevalent in observed galaxies. This difference compared to WHISP could be related to different spatial resolutions (the THINGS beam has a FWHM of $6\,"$). We point out that if central \HI holes are washed out in WHISP due to the lower resolution, then the same should happen to the TNG50 \HI maps as they are mock-observed at the WHISP resolution.} (see Figure~\ref{fig:ExampleImage} for a typical example). Central \HI holes could simultaneously explain the offsets in the concentration and $M_{20}$ statistics, with face-on galaxies deviating the most from observations as the impact of the central hole is maximized. In the following we discuss some possible sources of the apparently unrealistic prevalence of large central \HI holes in TNG50.

An obvious mechanism to create large central \HI holes is to ionize the neutral hydrogen or expel it from the galaxy center, for instance by feedback from active galactic nuclei (AGN). Indeed, \citet{Nelson2021} showed that related to AGN feedback, star-formation rates of TNG50 galaxies at $z\sim1$ are centrally suppressed, in agreement with observational data. In IllustrisTNG, AGN feedback occurs along a thermal and a kinetic mode, depending on the black hole accretion rate. For galaxies with stellar masses roughly below $10^{10.5}\,\mathrm{M}_\odot$, the AGN feedback is typically in the thermal mode with a continuous energy injection. For more massive galaxies, the kinetic mode kicks in and energy is injected in a pulsed, directed fashion into the surroundings (see \citealt{Weinberger2017} for more details). Furthermore, AGN radiation can ionize a substantial fraction of the hydrogen gas (see figure A6 of \citealt{Byrohl2021}). Recently, \citet{Ma2022} found that the kinetic AGN feedback in IllustrisTNG is very effective at redistributing neutral gas from the central to the outer galaxy regions compared to the SIMBA simulation, hinting that AGN feedback could indeed create central \HI holes. Upon examination of the neutral gas reservoirs of central and satellite galaxies in TNG100, \citet{Stevens2019} and \citet{Stevens2021} also speculate that some tension with observational data could arise due to AGN feedback ejecting/depleting neutral gas from the center.

As a first ad-hoc experiment to check if AGN feedback is responsible for the central \HI holes in TNG50, we exclude all galaxies that experienced AGN feedback in the kinetic mode ($\approx25\,\%$ of the TNG50 sample of 1130 galaxies). This slightly reduces the tension in the morphological statistics to $\Delta\tilde{C}\approx-0.23$ and $\Delta\tilde{M}_{20}\approx0.25$. If we instead discard all TNG50 galaxies with an above-average energy injection in the thermal mode, the tension reduces to $\Delta\tilde{C}\approx-0.18$ and $\Delta\tilde{M}_{20}\approx0.18$. While it seems plausible that AGN feedback is at least partially responsible for the tension in the concentration statistic, we caution that the AGN feedback properties of the TNG50 galaxies are probably cross-correlated with other galaxy properties (e.g. stellar mass, gas-to-stellar mass ratio) which could also influence the morphological statistics. To unequivocally pinpoint the impact of AGN feedback, different cosmological simulation runs varying the feedback parameters need to be considered, which is beyond the scope of this paper.

The impact of AGN feedback is also somewhat depending on the numerical mass resolution of the cosmological simulation. For the IllustrisTNG suite, the feedback parameters are calibrated for TNG100, and are then kept the same for all IllustrisTNG runs (\citealt{Pillepich2018}). Increasing the resolution of the simulation also increases the black hole accretion rate, especially in the thermal mode (see figure B1 of \citealt{Weinberger2017}). We test the impact of this numerical effect by repeating the construction of \HI maps and morphological analysis for TNG100, which has a mass resolution twenty times coarser than TNG50 (see Table~\ref{tab:TNGruns}). We find that the tension is reduced to $\Delta\tilde{C}\approx-0.16$ and $\Delta\tilde{M}_{20}\approx0.22$ when using TNG100.

Lastly, it is also possible that there is neutral hydrogen in galaxy centers but the gas is predominantly molecular. To assess this, we check if the choice of the \HI model affects the central \HI holes. We find almost no changes in the morphological statistics upon variation of the UV field model or partitioning scheme, except for the UV-independent partitioning scheme S14 (which increases the tension in the $C$ and $M_{20}$ statistics). However, when using `local' hydrogen partitioning schemes based on volume instead of surface densities (described in more detail in Section~\ref{sec:LocalSchemes}), we find that the tension reduces to $\Delta\tilde{C}\approx-0.19$ and $\Delta\tilde{M}_{20}\approx0.20$ (using the local GD14 scheme with the SKIRT UV field, this choice hardly affects the result). This could be related to the Jeans approximation breaking down for the dense galactic centers, as discussed in more detail in Section~\ref{sec:LocalSchemes}.

We point out that the unusual central \HI holes could also be due to the AGN feedback in IllustrisTNG being \textit{too weak}. If AGN activity would evacuate the cold gas from the entire galaxy instead of just the central region, the concentration statistics of such galaxies would increase and better match the WHISP data. It is also possible that such a reduction of cold gas in the galaxy leads to the object not being selected in the TNG50 sample, as it would be unobservable with WHISP. Indeed, results from \citet{Ma2022} indicate that for TNG100, the AGN feedback quenches the star formation but only redistributes the cold gas instead of decreasing the galactic cold gas reservoir as is observed (\citealt{Guo2021}). On the other hand, we caution that \citet{Ma2022} did not mimic the observational steps to mock-observe the simulated galaxies. \citet{Stevens2019} used the same observational data and hydrogen postprocessing method as \citet{Ma2022} but a careful mock-observation routine, finding excellent agreement for the cold gas reservoirs of TNG100 and observations.

The prevalence of central \HI holes is reminiscent of the finding of \citet{Bahe2016}, who discover unrealistic holes in the \HI disks of galaxies in the EAGLE simulation. \citet{Bahe2016} attribute this to the implementation of the stellar feedback in the physical model of EAGLE. Along similar lines, we partly attribute the large central \HI holes in TNG50 to the AGN feedback implementation. The effect is stronger in the higher-resolution simulation TNG50 compared to TNG100 due to the IllustrisTNG calibration scheme. \citet{Diemer2019} found results along similar lines as the \HI surface density profiles of IllustrisTNG galaxies exhibit central deficits compared to observational data from the Bluedisk survey (\citealt{Wang2014}), with larger mismatches for higher resolution simulations. However, we note that the usage of local partitioning schemes reduces the prevalence of central \HI holes. We conclude that these different effects (and potentially other systematics that we have not uncovered here) conspire to lead to the central \HI holes and eventually the mismatch in the $C$ and $M_{20}$ statistics.

\section{HI column density distribution function}\label{sec:HI CDDF}

The \HI column density distribution function is a widely used metric to quantify the distribution of atomic hydrogen in the Universe (e.g. \citealt{RyanWeber2003}; \citealt{Gabriel2011}; \citealt{Noterdaeme2012}; \citealt{Rahmati2015}). In this study, we use the \HI CDDF as the second metric (apart from the \HI morphologies) to compare TNG50 to observational data. The CDDF is insensitive to the spatial distribution of the H{\,\sc{i}}, it only probes \textit{how much} atomic hydrogen per column density interval exists in the galaxy population. Hence, this CDDF comparison is somewhat complementary to the analysis of \HI morphologies. We describe how we extract the \HI CDDF from TNG50 (Section~\ref{sec:CDDF_TNG}) and WHISP (Section~\ref{sec:CDDF_WHISP}), and compare them in Section~\ref{sec:CDDFresult}. We only discuss the default \HI model (GD14 + SKIRT UV field) for TNG50 in this section and compare to other partitioning schemes in Section~\ref{sec:LocalSchemes}.

\subsection{CDDF measurement from TNG50}\label{sec:CDDF_TNG}

As we intend to compare the \HI CDDF from TNG50 to observational data, we use the full mock map algorithm (Section~\ref{sec:MockMaps}) to generate 1130 synthetic \HI maps (in the three different angular resolutions of the WHISP survey) for TNG50 galaxies. We then follow the observational procedure to compute the \HI CDDF, which consists of summing the area covered by columns within a specific column density interval over all galaxies (e.g. \citealt{Zwaan2005}; \citealt{Szakacs2022}):

\begin{equation}\label{eq:CDDF}
    \mathrm{CDDF_{H\,\textsc{i}}}=\frac{c}{H_0}\frac{\sum_j\phi(M_\mathrm{H\,\textsc{i}}^j)\,w(M_\mathrm{H\,\textsc{i}}^j)\,A(N_\mathrm{H\,\textsc{i}})^j}{N_\mathrm{H\,\textsc{i}}\,\ln10\,\Delta(\log_{10}N_\mathrm{H\,\textsc{i}})},
\end{equation}
where $c$ denotes the speed of light, $N_\mathrm{H\,\textsc{i}}$ is the \HI column density, and $\Delta(\log_{10}N_\mathrm{H\,\textsc{i}})$ is the constant logarithmic column density bin spacing. This equation is applicable when the sample from which the CDDF is derived (in our case the 1130 mock-selected TNG50 galaxies) is drawn from a broader underlying galaxy population (the base TNG50 sample of 12431 galaxies). The CDDF of the Universe can then be computed by scaling the CDDF from the smaller sample using some galaxy property (e.g. stellar mass or \HI mass) in which the underlying broader sample is assumed to be complete\footnote{We remark that CDDFs calculated in this fashion are not necessarily representative of the broader galaxy population since only the $M_\mathrm{H\,\textsc{i}}$ distributions are matched. This does not affect our comparison between TNG50 and WHISP, as we follow the same methodology for the simulated and observed \HI{} CDDF. Hence, the mock TNG50/WHISP samples are equally (un-)representative of their broader galaxy populations.}. For the \HI CDDF, the \HI mass is a natural choice for this galaxy property (also adopted by e.g. \citealt{Zwaan2005}). Hence, the sum in Eqn.~\ref{eq:CDDF} runs over bins of \HI mass. The area function $A(N_\mathrm{H\,\textsc{i}})^j$ denotes the area covered by columns within the (logarithmic) column density interval summing over all galaxies (in the smaller sample) within the $j$-th \HI mass bin. The weight factor $w(M_\mathrm{HI}^j)$ corresponds to one over the number of galaxies in the smaller sample within the $j$-th \HI mass bin, thereby normalizing the area function. To scale the CDDF to the entire galaxy population, the normalized area function ($w(M_\mathrm{H\,\textsc{i}})A(N_\mathrm{H\,\textsc{i}})$) measured from the smaller sample is then multiplied with the abundance of objects in the broader sample in the $j$-th \HI mass bin per volume, $\phi(M_\mathrm{HI}^j)$. Hence, $\phi(M_\mathrm{H\,\textsc{i}})$ corresponds to the \HI mass function (derived from the broader sample), but without dividing by the (logarithmic) \HI mass bin width.

For our calculation of the \HI CDDF in TNG50, we compute the area function and the \HI masses for the weight factor from the 1130 mock \HI maps\footnote{Instead of computing the \HI masses from the mock \HI maps we could also use the `intrinsic' \HI masses of the galaxies, simply summing the \HI masses of the gas cells. However, this introduces inconsistencies with the observational procedure and the normalization of the CDDF.}. $\phi(M_\mathrm{H\,\textsc{i}})$ is calculated from the base TNG50 sample of 12431 galaxies. This approach is equivalent to observational determinations of the CDDF, where a smaller galaxy sample is observed with high spatial resolution to derive the area function. The area function is then scaled using the \HI mass function derived from blind galaxy surveys which measure the broader galaxy population. For cosmological simulations, we have the option to generate spatially resolved \HI maps directly for the broad galaxy population, circumventing the scaling of the CDDF from a smaller galaxy sample. As we want to emulate the observational procedure with realistic mock \HI maps (which requires a mock sample selection and putting the galaxies at the expected distances), we find the approach using Eqn.~\ref{eq:CDDF} more appropriate for this study. We test the simpler approach of measuring the CDDF from plain \HI maps of the broader galaxy population in Section~\ref{sec:CDDFcomparison}.

\subsection{CDDF measurement from WHISP}\label{sec:CDDF_WHISP}

\begin{figure*}
    \centering
    \includegraphics[width=\textwidth]{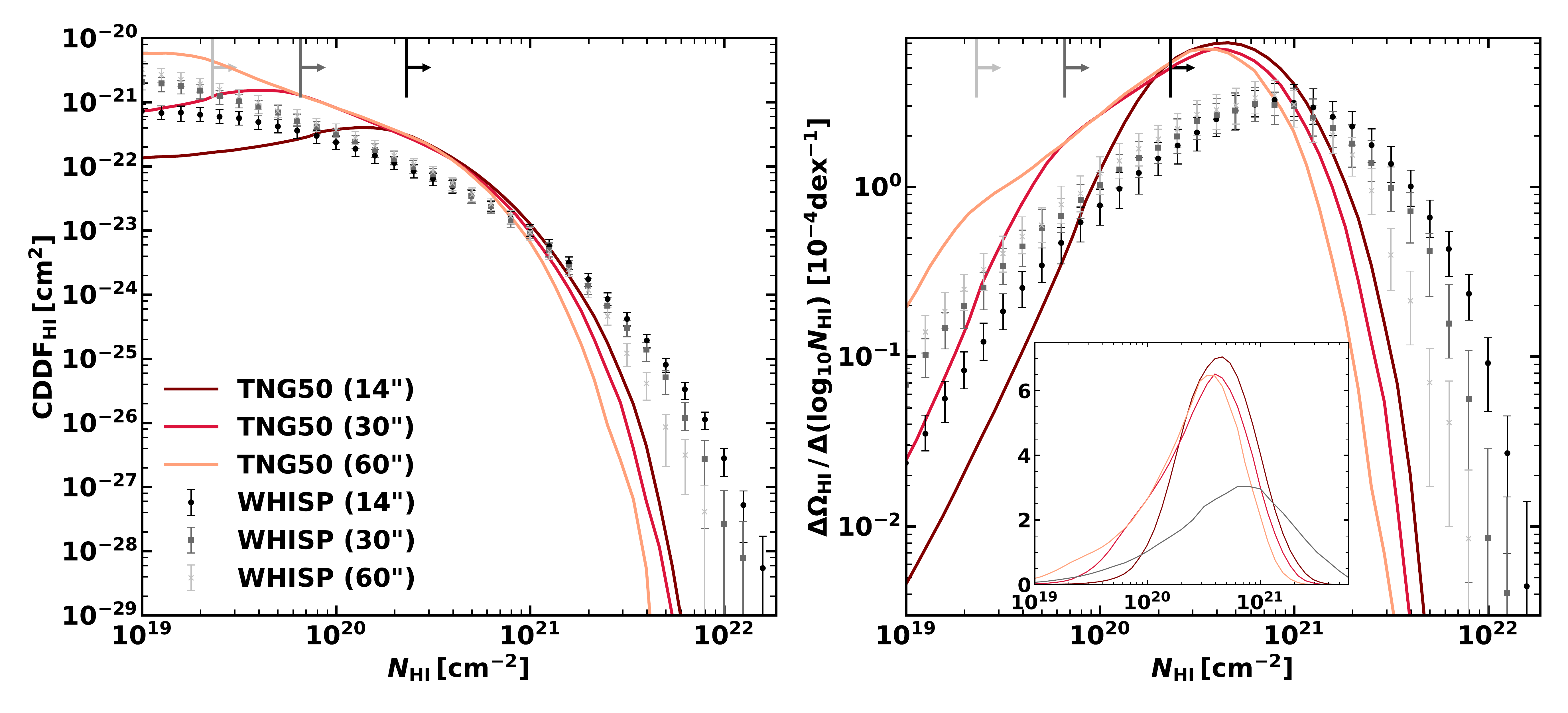}
    \caption{The \HI CDDF (left panel) and fractional contribution to $\Omega_\mathrm{H\,\textsc{i}}$ per column density interval (Eqn.~\ref{eq:CDDF_abundance}, right panel). The data points for the WHISP survey are taken from \citet{Zwaan2005} and corrected for various minor effects (Section~\ref{sec:CDDF_WHISP}). The colored lines show the results for TNG50 with varying angular resolutions, using the mock \HI maps and the default \HI model. Arrows at the top indicate the median 3-$\sigma$ sensitivity of the WHISP data (which equals the sensitivity of the TNG50 mock maps at the corresponding angular resolution). We show $\Delta\Omega_\mathrm{H\,\textsc{i}}/\Delta(\log_{10}N_\mathrm{H\,\textsc{i}})$ with a linear $y$-axis for the TNG50 results and the medium-resolution WHISP data in the inset. All curves are smoothed with a Gaussian filter with $\sigma\approx0.033\,\mathrm{dex}$.}
    \label{fig:CDDF}
\end{figure*}

The \HI CDDF at redshift zero can be measured with interferometric 21-cm surveys. A precise determination of the \HI CDDF is presented in \citet{Zwaan2005} based on WHISP data (see Section~\ref{sec:WHISPsurvey} for details about the WHISP survey). Although we could measure the \HI CDDF from the WHISP \HI maps ourselves, we refrain from doing so as the \citet{Zwaan2005} measurement constitutes the gold standard for the observational \HI CDDF at redshift zero and their result is used in many studies (e.g. \citealt{Braun2012}; \citealt{Rahmati2013a}; \citealt{Rahmati2013b}; \citealt{Szakacs2022}).

The \HI CDDF measurement from \citet{Zwaan2005} is based on the full WHISP survey (in its three available resolutions). Since the smaller WHISP sample of \citet{Naluminsa2021} that we use throughout this study is representative of the full WHISP sample, we do not expect that the usage of different WHISP samples affects our results. \citet{Zwaan2005} have scaled their measurement of the WHISP \HI CDDF to the broader galaxy population (see Eqn.~\ref{eq:CDDF}) using the \HI mass function from the blind HIPASS survey (\citealt{Zwaan2003}).

We add a cautionary note on $H_0$ correction factors. While the IllustrisTNG simulations were run with $h=0.6774$, the observational result from \citet{Zwaan2005} was obtained with $h=0.75$. It would be desirable to correct the observational result to the more recent value of $H_0$. For the \HI CDDF, however, this is an involved calculation due to the scaling factor $\phi$ (which is the HIMF modulo the mass bin width) in Eqn.~\ref{eq:CDDF}. This HIMF is parametrized as a Schechter function with parameters that themselves depend on the value of $H_0$. Hence, $H_0$ propagates in a non-polynomial way into the CDDF result from \citet{Zwaan2005}. We have run a test calculation of the \HI CDDF based on the WHISP \HI maps and the HIPASS mass function from \citet{Zwaan2003} to emulate the \citet{Zwaan2005} calculation (the actual CDDF calculation is more complex as \citealt{Zwaan2005} split the WHISP galaxies by morphological type and use type-specific mass functions). We find that updating $h$ from 0.75 to 0.6774 decreases the \HI CDDF by $\approx10-20\,\%$, except for the highest columns ($N_\mathrm{H\,\textsc{i}}>1.5\times10^{22}\,\mathrm{cm}^{-2}$) where the CDDF increases by $\approx10\,\%$. Since the $H_0$ correction factors that we find are much smaller than the difference between the original \citet{Zwaan2005} result and our test calculation, we keep using the \citet{Zwaan2005} result but multiply it with 0.85 to correct it to our value of $h=0.6774$.

Furthermore, the various biases in the HIMF from \citet{Zwaan2003} (e.g. selection effects, \HI self-absorption) are not taken into account by \citet{Zwaan2005} when calculating the WHISP CDDF. \citet{Zwaan2003} find that the cumulative effect of these biases reduces the total \HI abundance in the Universe by $11.6\,\%$. As the CDDF propagates linearly into the \HI abundance (see Eqn.~\ref{eq:CDDF_abundance}), we additionally decrease the WHISP CDDFs uniformly by $11.6\,\%$.

For studies of the \HI CDDF, which extends over nine orders of magnitude for $N_\mathrm{H\,\textsc{i}}\sim10^{19}-10^{22}\,\mathrm{cm}^{-2}$, these correction factors are completely negligible. For our study, we also consider the \HI abundance (which effectively corresponds to the normalization of the \HI CDDF), which scales linearly with any correction factors to the CDDF. Hence, the $H_0$ and HIMF correction factors are not negligible for our purpose.

\subsection{CDDF results}\label{sec:CDDFresult}

Since the \HI CDDF measures the area covered by columns per column density interval, the total \HI abundance in the Universe can be derived from the CDDF. Denoting the \HI abundance $\Omega_\mathrm{H\,\textsc{i}}$ as a fraction of the critical density of the Universe, $\Omega_\mathrm{H\,\textsc{i}}=\rho_\mathrm{H\,\textsc{i}}/\rho_c$ with $\rho_\mathrm{H\,\textsc{i}}$ the \HI mass density in the Universe and $\rho_c=1.274\times10^{11}\,\mathrm{M}_\odot\mathrm{Mpc}^{-3}$, we can write the fractional contribution to $\Omega_\mathrm{H\,\textsc{i}}$ per (logarithmic) column density interval as follows:

\begin{equation}\label{eq:CDDF_abundance}
\begin{split}
    \frac{\Delta\Omega_\mathrm{H\,\textsc{i}}}{\Delta(\log_{10}N_\mathrm{H\,\textsc{i}})}&=\frac{m_\mathrm{H\,\textsc{i}}}{\rho_cV}\frac{N_\mathrm{H\,\textsc{i}}\sum_j\phi(M_\mathrm{H\,\textsc{i}}^j)\,w(M_\mathrm{H\,\textsc{i}}^j)\,A(N_\mathrm{H\,\textsc{i}})^j}{\Delta\bigl(\log_{10}N_\mathrm{H\,\textsc{i}}\bigr)} \\ &= \frac{m_\mathrm{H\,\textsc{i}}H_0\ln(10)}{\rho_cc}\bigl(N_\mathrm{H\,\textsc{i}}\bigr)^2\cdot\mathrm{CDDF_{H\,\textsc{i}}}(N_\mathrm{H\,\textsc{i}}),
\end{split}
\end{equation}
where $m_\mathrm{H\,\textsc{i}}$ is the mass of a hydrogen atom, $V$ is the volume of the survey or the simulation box, and the sum in the first line of Eqn.~\ref{eq:CDDF_abundance} runs over bins of \HI mass.

We show our main CDDF result, Figure~\ref{fig:CDDF}, as the conventional \HI CDDF (left panel) and as fractional contribution to $\Omega_\mathrm{H\,\textsc{i}}$ (right panel). Observational results from \citet{Zwaan2005} (slightly modified as described in Section~\ref{sec:CDDF_WHISP}) are shown in grey to black markers for the three available WHISP resolutions. The TNG50 CDDFs for the different angular resolutions are displayed as colored lines. For our comparison it is more convenient to analyze $\Delta\Omega_\mathrm{H\,\textsc{i}}/\Delta(\log_{10}N_\mathrm{H\,\textsc{i}})$ instead of the conventional CDDF for two reasons: $\Delta\Omega_\mathrm{H\,\textsc{i}}/\Delta(\log_{10}N_\mathrm{H\,\textsc{i}})$ spans much fewer orders of magnitude which enables a more precise visual comparison for the different column density distributions. Furthermore, it also relates directly to the \HI abundance as $\Omega_\mathrm{H\,\textsc{i}}$ effectively corresponds to the integral of $\Delta\Omega_\mathrm{H\,\textsc{i}}/\Delta(\log_{10}N_\mathrm{H\,\textsc{i}})$ over the column density. To visualize this integral we also show $\Delta\Omega_\mathrm{H\,\textsc{i}}/\Delta(\log_{10}N_\mathrm{H\,\textsc{i}})$ with a linear $y$-axis in the inset in Figure~\ref{fig:CDDF}, where the areas under the curves directly correspond to the \HI abundance of TNG50 or the measured \HI abundance in the Universe.

We caution that the comparison of the column density distributions becomes unreliable below the sensitivity limits, indicated by the arrows at the top of Figure~\ref{fig:CDDF}. The segmentation for our TNG50 \HI maps is more aggressive (the main object cutout is smaller) than the 3-$\sigma$-clipped WHISP maps used by \citet{Zwaan2005} such that the TNG50 CDDFs decrease quickly for column densities below the sensitivity limits (note that the median sensitivity limits of WHISP and the TNG50 mock maps are equal due to our mock map generation algorithm, see Section~\ref{sec:MockMaps}).

Intermediate column densities ($N_\mathrm{H\,\textsc{i}}\approx10^{20}-10^{21}\,\mathrm{cm}^{-2}$) are significantly more abundant in TNG50 than in WHISP. This leads to a substantial mismatch in the \HI abundance comparing TNG50 and observational data. We point out that the \HI abundance (equivalent to the normalization of the CDDF) as visualized in the inset of Figure~\ref{fig:CDDF} is actually completely independent of the WHISP data or the TNG50 \HI mock maps due to the scaling factor $\phi(M_\mathrm{H\,\textsc{i}})$ in Eqn.~\ref{eq:CDDF}. Instead, the normalizations of the CDDFs are given by $\Omega_\mathrm{H\,\textsc{i}}$ as measured by the blind HIPASS survey (for the WHISP CDDFs) or the total amount of \HI in the TNG50 base galaxy sample of 12431 galaxies (for the TNG50 CDDFs), respectively. Hence, the tension in the CDDF at these intermediate column densities points towards an excess of \HI in TNG50 or an underestimation of $\Omega_\mathrm{H\,\textsc{i}}$ in blind \HI surveys, which we discuss in more detail in Section~\ref{sec:HIabundance}.

At the highest column densities ($N_\mathrm{H\,\textsc{i}}>10^{21}\,\mathrm{cm}^{-2}$), the effect of beam smearing becomes substantial both for WHISP and TNG50, which manifests in smearing out the high-density peaks. In this column density range, TNG50 significantly underpredicts the CDDF compared to WHISP. We find that there is neutral hydrogen in TNG50 in these column densities (see also \citealt{Szakacs2022}), but the hydrogen partitioning that we perform is very effective at converting \HI into H$_2$ such that the abundance of high \HI column densities in the simulation is lower than observed. We discuss this tension and a potential resolution in more detail in Section~\ref{sec:LocalSchemes}. Lastly, we point out that our result contrasts with other studies of the \HI CDDF (\citealt{Rahmati2013a}; \citealt{Rahmati2013b}; \citealt{Villaescusa2018}) who find a good agreement between WHISP and cosmological simulations, especially at the high-column end. We attribute this primarily to the beam smoothing of the mock \HI maps, which is not performed in these studies (see Section~\ref{sec:CDDFcomparison} for more details).

\section{Discussion}\label{sec:Discussion}

\subsection{Normalization of the CDDF: HI abundance in IllustrisTNG}\label{sec:HIabundance}

\begin{figure*}
    \centering
    \includegraphics[width=\textwidth]{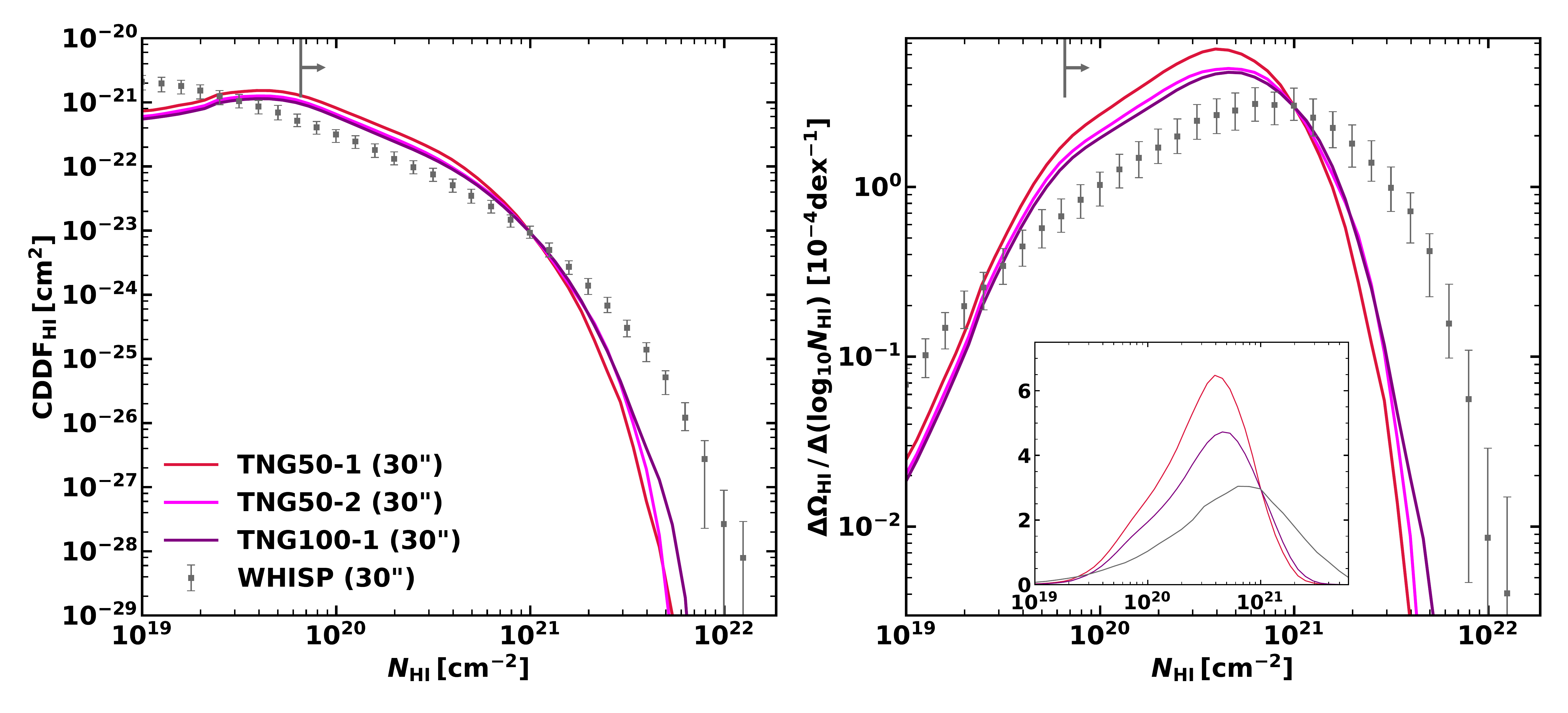}
    \caption{Same as Figure~\ref{fig:CDDF}, varying the IllustrisTNG simulations instead of the angular resolution of the simulation data. The inset in the right panel shows just the TNG50-1, TNG100-1, and medium-resolution WHISP result.}
    \label{fig:CDDF_VarSim}
\end{figure*}

As shown in the inset of Figure~\ref{fig:CDDF}, the CDDF normalization (equivalent to $\Omega_\mathrm{H\,\textsc{i}}$) in TNG50 significantly exceeds the observational measurement. The observational value of $\Omega_\mathrm{H\,\textsc{i}}$ at redshift zero is most reliably determined with blind \HI surveys (see figure 14 of \citealt{Rhee2018} for a data compilation). For HIPASS, \citet{Zwaan2005b} find $\Omega_\mathrm{H\,\textsc{i}}=(3.9\pm0.4^\mathrm{stat}\pm0.4^\mathrm{sys})\times10^{-4}$, consistent with the result for ALFALFA of $\Omega_\mathrm{H\,\textsc{i}}=(4.0\pm0.1^\mathrm{stat}\pm0.6^\mathrm{sys})\times10^{-4}$ (\citealt{Jones2018}). Both of these measurements are corrected for various biases such as selection bias or \HI self-absorption. Furthermore, we corrected these \HI abundances to our adopted value of the Hubble constant, $h=0.6774$.

As the WHISP CDDF is scaled with the HIPASS HIMF, we expect the exactly same value of $\Omega_\mathrm{H\,\textsc{i}}=3.9\times10^{-4}$ when calculating the \HI abundance from the WHISP CDDF. With the raw CDDF from \citet{Zwaan2005} we obtain $\Omega_\mathrm{H\,\textsc{i}}=5.1\times10^{-4}$, but with the Hubble and HIMF biases correction factors (see Section~\ref{sec:CDDF_WHISP}) the \HI abundance is reduced to the expected value of $\Omega_\mathrm{H\,\textsc{i}}=3.9\times10^{-4}$ (with very little impact on the WHISP resolution). Hence, we consider our correction factors somewhat realistic, at least at the level of total \HI abundance.

Studies investigating the \HI abundance in IllustrisTNG (\citealt{Villaescusa2018}; \citealt{Diemer2019}, \citealt{Yates2021}) found higher values of $\Omega_\mathrm{H\,\textsc{i}}$ compared to blind \HI surveys. The magnitude of this discrepancy depends on the simulation version of the IllustrisTNG suite and the postprocessing routines. Summing all \HI in galaxies in the TNG100 simulation, \citet{Diemer2019} report $\Omega_\mathrm{H\,\textsc{i}}=(5.4\pm0.5)\times10^{-4}$, in tension with the observational results. For TNG50, \citet{Diemer2019} report $\Omega_\mathrm{H\,\textsc{i}}=(7.2\pm1.1)\times10^{-4}$, in significant excess of the data. The error bars on these values correspond to the scatter due to the different partitioning schemes considered by \citet{Diemer2019}. Using our default \HI model (GD14 plus SKIRT UV field) for our base sample of 12431 galaxies in TNG50, we obtain $\Omega_\mathrm{H\,\textsc{i}}= 6.0\times10^{-4}$. This is at the lower edge of the value reported by \citet{Diemer2019}, since the GD14 scheme typically predicts the highest molecular fractions (together with the S14 scheme) and because we miss a few gas-rich galaxies with very low stellar mass due to our galaxy sample selection.

\subsubsection{Observational considerations}

A natural explanation for this excess of \HI in IllustrisTNG is that $\Omega_\mathrm{H\,\textsc{i}}$ is underestimated by the observations. \HI gas in low-column-density gas could be difficult to detect, while very high columns are prone to \HI self-absorption and hence an underestimation of the actual \HI column density. Indeed, \citet{Braun2012} find based on observations of M31, M33 and the LMC that self-absorption hides a significant fraction of \HI gas. Correcting for this self-absorption, \citet{Braun2012} claim that $\Omega_\mathrm{H\,\textsc{i}}=(5.9\pm0.9)\times10^{-4}$. However, we remark that this study is questioned by more recent findings of \citet{Koch2021} who use a more thorough \HI model to measure the self-absorption correction. In any case, the distribution of \HI in terms of column densities provides a convenient means to explore if we miss \HI gas when measuring $\Omega_\mathrm{H\,\textsc{i}}$ from 21-cm emission data.

From the linear inset of Figure~\ref{fig:CDDF}, we learn that $\Omega_\mathrm{H\,\textsc{i}}$ is dominated by an intermediate column density regime of $N_\mathrm{H\,\textsc{i}}\approx10^{19}-4\times10^{21}\,\mathrm{cm}^{-2}$. From our mock \HI maps it is impossible to assess if \HI column densities below $N_\mathrm{H\,\textsc{i}}\approx10^{19}$ contribute substantially to $\Omega_\mathrm{H\,\textsc{i}}$ as the mock \HI maps are segmented at approximately this column density (for the low-resolution maps). Generating plain \HI maps (without noise and segmentation) for the full TNG50 base sample (see Section~\ref{sec:CDDFcomparison} for more details), we find that column densities in the range $10^{17}-10^{19}\,\mathrm{cm}^{-2}$ only contribute 2.2\,\% to the total \HI abundance. At the other extreme, column densities above $\approx4\times10^{21}\,\mathrm{cm}^{-2}$ hardly contribute to $\Omega_\mathrm{H\,\textsc{i}}$ both for TNG50 and the WHISP data, indicating that \HI self-absorption does not hide a significant amount of \HI gas (\citealt{Braun2012} find that \HI self-absorption only becomes important above $N_\mathrm{H\,\textsc{i}}\approx10^{22}\,\mathrm{cm}^{-2}$).

It is also possible that blind \HI surveys simply miss a significant population of objects. If the additional \HI gas in TNG50 stemmed from galaxies with very small \HI masses, these objects could be missed, while very heavy galaxies could be rare and hence difficult to observe. For TNG50, we find from the HIMF (Figure~\ref{fig:MassFunction}) that the dominating objects for $\Omega_\mathrm{H\,\textsc{i}}$ have \HI masses within $3\times10^8-3\times10^{10}\,\mathrm{M}_\odot$. We do not expect that our base sample selection (imposing $M_\mathrm{gas}>10^7\,\mathrm{M}_\odot$ and $M_\star>10^7\,\mathrm{M}_\odot$) or the limited simulation volume affects this finding. Galaxies in this mass range are easy to detect with blind \HI surveys, for ALFALFA the sensitivity is good enough such that the HIMF can be precisely determined for \HI masses larger than $10^7\,\mathrm{M}_\odot$ (figure 2 of \citealt{Jones2018}). We conclude that the mismatch in $\Omega_\mathrm{H\,\textsc{i}}$ (i.e. the CDDF normalization) between TNG50 and WHISP is due to an excess of atomic hydrogen in the simulation and not an underestimation by the observational data.

\subsubsection{Simulation considerations}

\begin{table}
    \centering
    \begin{tabular}{ccc}
         Data set & $\Omega_\mathrm{H\,\textsc{i}}$ [$10^{-4}$] & Reference \\ \hline
         ALFALFA & $4.0\pm0.1^\mathrm{stat}\pm0.6^\mathrm{sys}$ & \citet{Jones2018} \\
         HIPASS & $3.9\pm0.4^\mathrm{stat}\pm0.4^\mathrm{sys}$ & \citet{Zwaan2005b} \\
         Local \HI CDDF  & $5.9\pm0.9$ & \citet{Rhee2018} \\ \hline
         TNG50-1 & $7.2\pm1.1$ & \citet{Diemer2019} \\
         TNG100-1 & $5.4\pm0.5$ & \citet{Diemer2019} \\
         TNG50-1 GD14 (aperture) & 6.0 (5.4) & This work \\
         TNG50-2  GD14 (aperture) & 5.1 (4.6) & This work \\
         TNG100-1 GD14 (aperture) & 4.8 (4.3) & This work \\
         \end{tabular}
    \caption{All values of the \HI abundance $\Omega_\mathrm{H\,\textsc{i}}$ that we consider in this study. The upper three entries are observational determinations of $\Omega_\mathrm{H\,\textsc{i}}$, the lower five entries correspond to results from IllustrisTNG. For the \citet{Diemer2019} values, the error bars correspond to the spread between different partitioning schemes. For the results from this work, we calculate the \HI abundance using only our default \HI model. The values inside the brackets denote results when correcting for the ALFALFA beam.}
    \label{tab:HIabundance}
\end{table}

A possible resolution of this $\Omega_\mathrm{H\,\textsc{i}}$ discrepancy consists in mimicking the blind \HI surveys like HIPASS or ALFALFA in IllustrisTNG, taking sensitivity limits and other observational effects into account (which were neglected in previous studies of $\Omega_\mathrm{H\,\textsc{i}}$ in IllustrisTNG). While it is beyond the scope of this paper to fully emulate the ALFALFA or HIPASS observational setup, we assess the impact of the observational effect which probably affects the inferred \HI abundance most, which is beam smoothing. An ALFALFA-like beam smoothing was used in \citet{Diemer2019} (based on \citealt{Bahe2016}) to emulate the \HI mass function measurement in IllustrisTNG. We follow their approach and apply a Gaussian beam with $\sigma=70\,\mathrm{kpc}$, i.e. we multiply the \HI mass of each gas cell by $\exp[-r^2/(2\sigma^2)]$, with $r$ being the 2D distance of the gas cell to the galaxy center in face-on projection (see (i) in \ref{sec:PlainMaps} for the algorithm to rotate galaxies into face-on projection). For our default \HI model, we find that applying this beam smoothing effect reduces the \HI abundance in TNG50 by ten percent to $\Omega_\mathrm{H\,\textsc{i}}=5.4\times10^{-4}$. Hence, mimicking blind \HI surveys reduces the tension in \HI abundance and the CDDF normalization, but does not resolve it for TNG50.

The IllustrisTNG simulation resolution also affects the \HI abundance. For the base galaxy samples of TNG50-2 (TNG100) and the default \HI model, we find $\Omega_\mathrm{H\,\textsc{i}}=5.1\times10^{-4}$ ($\Omega_\mathrm{H\,\textsc{i}}=4.8\times10^{-4}$). Applying the ALFALFA-like beam smoothing further reduces these values by approximately ten percent. All observational and simulated inferences of $\Omega_\mathrm{H\,\textsc{i}}$ are summarized in Table~\ref{tab:HIabundance}. The \HI abundances of the lower-resolution simulations TNG50-2 and TNG100, taking the ALFALFA beam into account, are consistent with the observational values.

We show the \HI CDDF for various IllustrisTNG simulations (calculated as described in Section~\ref{sec:CDDF_TNG}) in Figure~\ref{fig:CDDF_VarSim}. Note that the scaling factor $\phi(M_{H\,\textsc{i}})$ was computed as in Section~\ref{sec:CDDF_TNG}, i.e. without ALFALFA-like beam smoothing. This means that the normalization of the IllustrisTNG CDDFs corresponds to the larger values in Table~\ref{tab:HIabundance} (e.g. $4.8\times10^{-4}$ for TNG100). Differences in the \HI CDDF (left panel) for the various IllustrisTNG runs appear marginal, except at the highest column densities which are dominated by very few high-column-density pixels. In the linear inset in the right panel, it becomes clear that TNG100 is closer to the observational data than TNG50, which is due to the lower \HI abundance in TNG100. Furthermore, we note that the effect of varying the simulation volume is negligible compared to varying the simulation resolution, as TNG100 and TNG50-2 have similar \HI CDDFs (the TNG100 volume is roughly eight times larger than the TNG50-2 one, while the TNG100 mass resolution is comparable to TNG50-2, see Table~\ref{tab:TNGruns}).

The $\Omega_\mathrm{H\,\textsc{i}}$ excess in TNG50 is reminiscent of the luminosity functions analyzed in \citet{Trcka2022}, who find that the lower-resolution TNG50-2 provides a substantially better match to observational data than TNG50. This is expected given that the physical model calibration for IllustrisTNG was undertaken at the resolution of roughly TNG100 (TNG50-2 has a comparable resolution). Indeed, \citet{Pillepich2018} show that the stellar mass function in IllustrisTNG is converging, but not fully converged, as a function of numerical resolution. For a given dark matter halo, galaxies form slightly more stars at higher resolutions, shifting the stellar mass function. It is not surprising that these higher stellar masses lead to higher \HI masses\footnote{This argument assumes that the H\,\textsc{i}-to-stellar mass ratio is independent of numerical resolution. Indeed, \citet{Diemer2019} find that this is true to first order and the \HI fractions of various IllustrisTNG simulations agree well with an observational compilation from \citet{Calette2018}.} and ultimately too much atomic hydrogen in the TNG50 galaxy population.

\subsection{Hydrogen partitioning based on volume densities}\label{sec:LocalSchemes}

\begin{figure*}
    \centering
    \includegraphics[width=\textwidth]{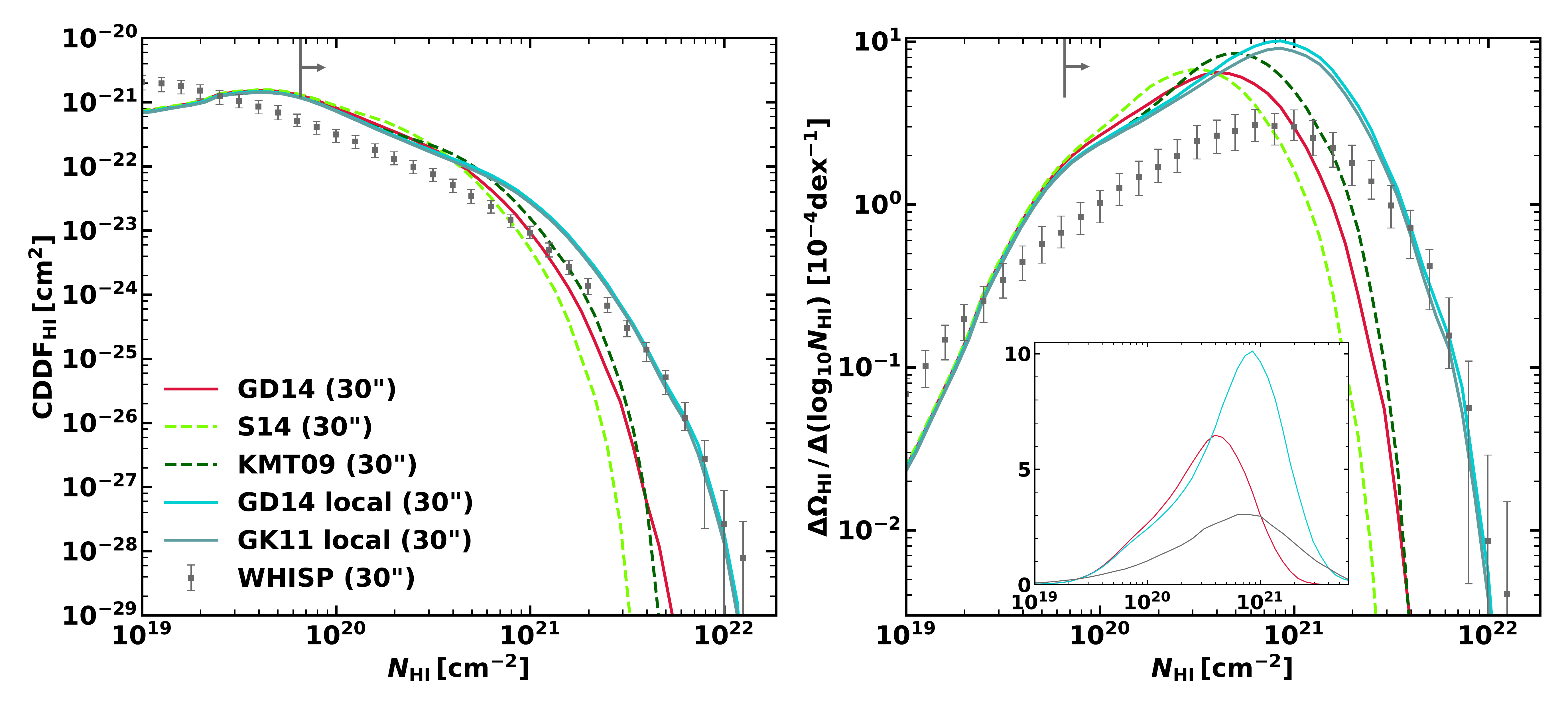}
    \caption{Same as Figure~\ref{fig:CDDF}, varying the hydrogen partitioning scheme instead of the angular resolution for the simulation data. The default \HI model (GD14 with the SKIRT UV field) is shown in red, dashed lines indicate the UV-independent partitioning schemes S14 and KMT09. Furthermore, the UV-dependent local partitioning schemes of \citet{Gnedin2011} (GK11 local) and \citet{Gnedin2014} (GD14 local) are shown in cyan. We have only used the SKIRT UV field for the application of the local partitioning schemes. The GD14, GD14 local, and medium-resolution WHISP results are also displayed in the linear inset.}
    \label{fig:CDDFmodels}
\end{figure*}

The simulation-calibrated partitioning schemes of GK11 and GD14 also exist in `local' variants that are based on volume instead of surface densities. At first sight this is a very attractive feature because it completely circumvents the difficulties of assigning surface densities to gas cells in simulations. According to \citet{Gnedin2011}, a cosmological simulation using their local partitioning scheme needs to have a spatial resolution of $L_\mathrm{cell}\sim100\,\mathrm{pc}$. Additionally, \citet{Gnedin2014} introduce a correction factor for the spatial resolution which is tested up to $L_\mathrm{cell}\sim500\,\mathrm{pc}$, well in excess of the median size of star-forming gas cells in TNG50 of 138\,pc (\citealt{Nelson2019dataRelease}).

For the \HI morphologies, we find that the usage of local partitioning schemes provides a slightly better match to the concentration and $M_{20}$ indices between TNG50 and WHISP galaxy samples. Here, we test the local partitioning schemes in terms of the \HI CDDF in Figure~\ref{fig:CDDFmodels}. The application of these schemes to IllustrisTNG with the relevant equations is described in Appendix~\ref{sec:GK11 local} and~\ref{sec:GD14 local}. Since the predictions of the local partitioning models are similar upon variation of the UV field scheme we only consider the SKIRT UV field here. For completeness we also check the \HI CDDF using the two UV-independent partitioning schemes (KMT09 and S14) in Figure~\ref{fig:CDDFmodels}. We find that the KMT09 model (as well as the UV-dependent GK11 and K13 schemes which are not shown here) produces results comparable to the default GD14 partitioning scheme, while S14 underestimates the abundance of high-column \HI gas even more than the other models.

The local partitioning schemes provide a perfect match at the high-column end, which is a substantial improvement over all other \HI models. The higher concentration values of the \HI maps using local partitioning schemes are related to this, as the central \HI holes are less prevalent and more \HI is formed in the dense galactic central regions (see Section~\ref{sec:Holes}). However, the \HI abundance is very high with the local partitioning schemes (visible in the linear inset in the right panel of Figure~\ref{fig:CDDFmodels}), and reaches almost $\Omega_\mathrm{H\,\textsc{i}}=10\times10^{-4}$ for TNG50, 2.5 times more than what is observed. This excess of \HI using local partitioning schemes persists to TNG100, where we find $\Omega_\mathrm{H\,\textsc{i}}=6.9\times10^{-4}$. This provides an a posteriori justification for refraining from using the local partitioning schemes, albeit they are conceptually appealing as they circumvent the need to estimate column densities for the 3D cosmological simulations.

A naive explanation for the difference between conventional, column-density based and local partitioning schemes is that the Jeans approximation breaks down at high column densities. For the presumably star-forming gas cells in these high-density regions, the internal energy of the gas cell which is recorded by IllustrisTNG is an average over cold and hot phases in the context of the two-phase model of \citet{Springel2003}. This means that the internal energy of the cold, neutral gas phase is overestimated, artificially increasing the Jeans length (see Eqn.~\ref{eq:Jeans}) for the neutral star-forming gas. This unrealistically boosts neutral hydrogen column densities estimated from the Jeans approximation and molecular hydrogen formation, thereby explaining the low abundance of \HI in high-density regions. \citet{Diemer2018} tested this issue by computing neutral hydrogen surface densities with the Jeans approximation, both with the internal energy recorded by IllustrisTNG and the internal energy of the cold gas (using $T=1000\,\mathrm{K}$ for the cold gas). Surprisingly, they find that the Jeans approximation with the internal energy recorded by IllustrisTNG fits the true surface densities (estimated from projecting the gas cells) better.

Instead of tweaking the Jeans approximation, a resolution for discrepancies in the morphological statistics and the CDDF is to use local partitioning schemes, under the assumption that there is simply too much neutral hydrogen in IllustrisTNG. For local partitioning schemes applied to TNG100, the \HI abundance is approximately 1.75 times higher than the observational value. Scaling down the \HI CDDF (computed from local partitioning schemes) by this factor yields an excellent agreement to the WHISP data, even at the high-column-density end. If the local partitioning schemes realistically split the gas into atomic and molecular phases, then the only way to reconcile the observational data with IllustrisTNG is to reduce the amount of neutral hydrogen in the simulation. To test if IllustrisTNG galaxies exhibit too much neutral gas requires mimicking both \HI and H$_2$ abundance measurements using consistent postprocessing methods (i.e. a consistent hydrogen partitioning scheme). While this is beyond the scope of this paper, we note that results from \citet{Popping2019} for TNG100 tentatively indicate that there is indeed an overabundance of molecular hydrogen (using conventional, column-density based partitioning schemes). It would be interesting to test if this H$_2$ overabundance persists when using local partitioning schemes.

\subsection{TNG-intrinsic CDDF and comparison to other studies}\label{sec:CDDFcomparison}

\subsubsection{CDDF from plain \HI maps}

\begin{figure*}
    \centering
    \includegraphics[width=\textwidth]{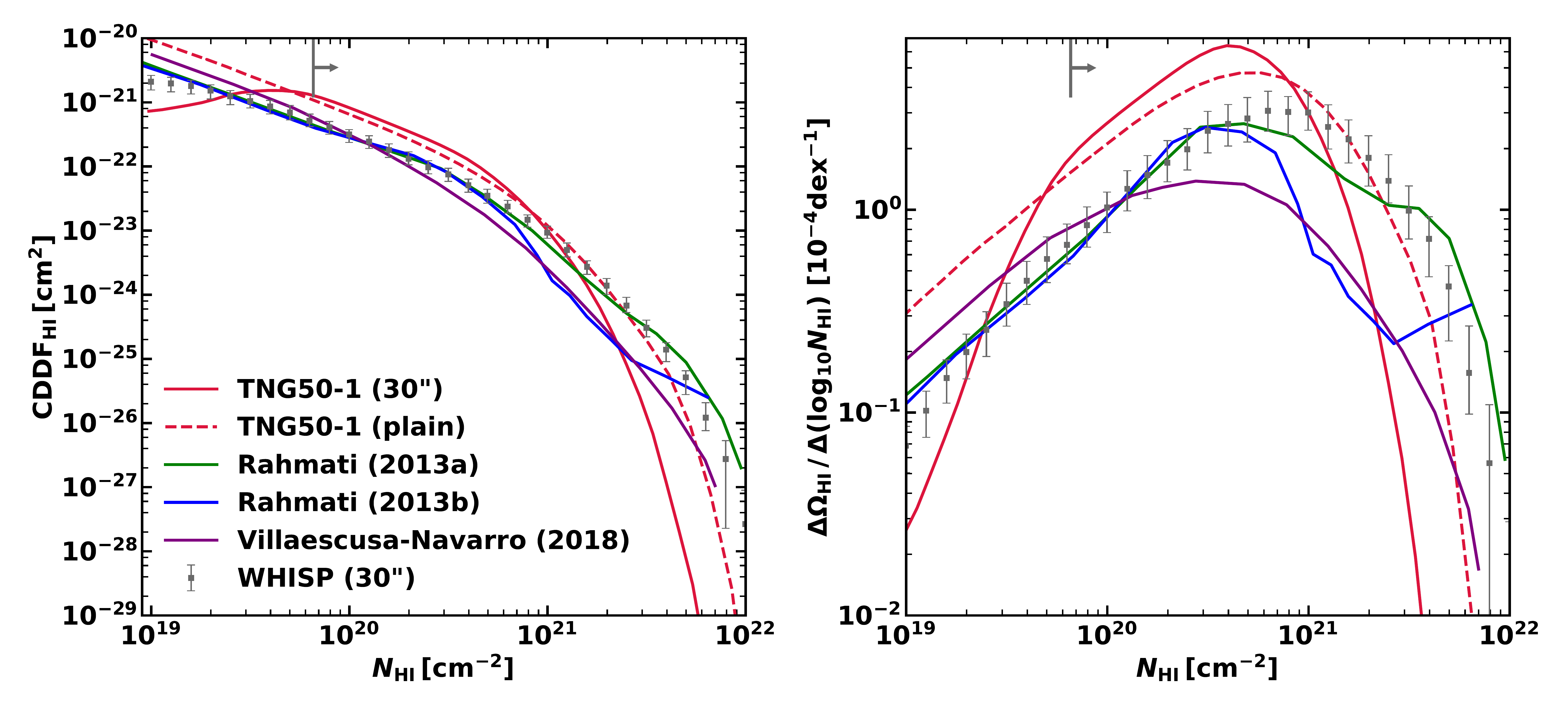}
    \caption{Same as Figure~\ref{fig:CDDF}, with TNG50-1 (30") corresponding to the CDDF calculated from the mock \HI maps (Section~\ref{sec:MockMaps}) using Eqn.~\ref{eq:CDDF}. The dashed red line indicates the TNG50-1 CDDF calculated from plain \HI maps (Section~\ref{sec:PlainMaps}) using Eqn.~\ref{eq:CDDF_plain}. Simulation results for the \HI CDDF at redshift zero from \citet{Rahmati2013a}, \citet{Rahmati2013b}, and \citet{Villaescusa2018} are also shown.}
    \label{fig:CDDFplain}
\end{figure*}

Our results for the \HI CDDF (Figure~\ref{fig:CDDF}), especially the mismatch between simulations and observations for high column densities, contrasts with similar studies of the CDDF (\citealt{Rahmati2013a}; \citealt{Rahmati2013b}; \citealt{Villaescusa2018}). We attribute the discrepancies to our mock \HI map creation (see Section~\ref{sec:MockMaps}) which takes observational effects into account, while other studies of the \HI CDDF in cosmological simulations adopted simpler prescriptions to generate synthetic \HI maps. To verify this and to assess the impact of the observational effects modelled in the context of our mock \HI map algorithm, we construct the \HI CDDF for TNG50 using simpler plain \HI maps (see Section~\ref{sec:PlainMaps}) in this section.

We generate plain \HI maps for the TNG50 base sample of 12431 galaxies, and calculate the CDDF directly from this broad sample without the need of a scaling factor as in Eqn.~\ref{eq:CDDF} (`TNG-intrinsic CDDF'):

\begin{equation}\label{eq:CDDF_plain}
    \mathrm{CDDF_{H\,\textsc{i}}}=\frac{c}{H_0}\frac{\sum_jA(N_\mathrm{H\,\textsc{i}})_j}{N_\mathrm{H\,\textsc{i}}\,\ln10\,\Delta(\log_{10}N_\mathrm{H\,\textsc{i}})\,V},
\end{equation}
where the sum runs over all galaxies in the (broad) sample. Typically, the approach to extract the CDDF from cosmological simulations is to project the full simulation box onto a 2D grid and measure the CDDF there. However, since we focus on the high-density end of the column density distribution which traces gas within galaxies, we find Eqn.~\ref{eq:CDDF_plain} more appropriate here.

The intrinsic CDDF generated from plain \HI maps is shown in Figure~\ref{fig:CDDFplain} (dashed red line). The beam smoothing effect on the mock \HI maps is clearly visible in Figure~\ref{fig:CDDFplain}, as the CDDF from the mock \HI maps (solid red line) has the highest column density peaks smoothed out and moved to intermediate densities, thereby increasing the tension with the WHISP CDDF. In fact, the CDDFs generated from plain \HI maps (especially from the lower-resolution TNG50-2 and TNG100 runs which are not shown in Figure~\ref{fig:CDDFplain}) show a good agreement with the observational data, as found in previous studies of the \HI CDDF. We have verified that the choice of the map resolution of the plain \HI maps ($128\times128$ by default) does not impact our CDDF result.

\subsubsection{Other studies of the CDDF in cosmological simulations}

A few studies which analyzed CDDFs in cosmological simulations exist. Here, we compare our CDDF to their findings to provide some additional context for our results. 

\citet{Rahmati2013a} computed the \HI CDDF for a set of cosmological simulations which explicitly track collisional ionization, radiation from hydrogen recombination and the UV background which sets the hydrogen ionization state. The CDDF itself is calculated by projecting the entire simulation box onto one side, contrary to our approach of summing up the column density distributions of individual galaxies. Molecular hydrogen formation is modelled with the empirical partitioning scheme of \citet{Blitz2006}. \citet{Rahmati2013a} find an excellent agreement at redshift zero between their \HI CDDF (green line in Figure~\ref{fig:CDDFplain}) and the WHISP measurement by \citet{Zwaan2005}, as their CDDF at the low-column end is lower and hence more in line with observational data. This difference probably arises due to the usage of different cosmological simulations (for instance, the reference simulation of \citealt{Rahmati2013a} has baryonic particle masses two orders of magnitude larger than TNG50). 

With radiation from local stellar sources added in \citet{Rahmati2013b}, the \HI CDDF sharply falls by 0.6 dex for $N_\mathrm{H\,\textsc{i}}>10^{21}\,\mathrm{cm}^{-2}$ due to a larger ionized hydrogen fraction at large column densities (blue line in Figure~\ref{fig:CDDFplain}). This leads to a significant underestimation of the \HI CDDF at these high columns. H$_2$ formation is not considered in \citet{Rahmati2013b}, leading to the upturn of the \HI CDDF at the highest column densities. We remark that the calculation of the neutral hydrogen fraction in IllustrisTNG is based on \citet{Rahmati2013a}, without the correction of local stellar sources. Since we base the calculation of the neutral hydrogen fraction in star-forming gas cells on the fraction of gas in the cold phase instead of the IllustrisTNG output, it is not directly clear how the inclusion of local ionizing sources would affect the \HI CDDF.

More recently, \citet{Villaescusa2018} postprocessed the TNG100 simulation with the UV-independent partioning scheme of KMT09 and computed the \HI CDDF in a similar fashion as \citet{Rahmati2013a}, i.e. by projecting the full simulation box onto one side with $20`000\times20`000$ points. A notable difference to our method is the usage of the cell size instead of the Jeans length for the calculation of surface densities which are used for the KMT09 partitioning scheme. We find that for star-forming gas cells, the cell sizes are smaller than the Jeans lengths by approximately one order of magnitude. This should significantly lower the molecular fractions for the gas cells. Furthermore, their calculation of the CDDF consists of modelling the gas cells as uniform density spheres, and computing the column density from the length of the line-of-sight segment intersecting the spheres.

We find that their \HI CDDF (purple line in Figure~\ref{fig:CDDFplain}) is lower than ours for all columns below $\approx5\times10^{21}\,\mathrm{cm}^{-2}$. The discrepancy rises to almost one order of magnitude at $N_\mathrm{H\,\textsc{i}}\approx10^{21}\,\mathrm{cm}^{-2}$, and this result holds even when we apply the KMT09 partitioning scheme to the TNG100 simulation for our result. We attribute this discrepancy to the usage of uniform density spheres in \citet{Villaescusa2018} when calculating the \HI CDDF. We find that for $N_\mathrm{H\,\textsc{i}}\sim10^{21}\,\mathrm{cm}^{-2}$, the radius of a gas cell (modelled as a sphere) does not exceed $R_\mathrm{sphere}=290\,\mathrm{pc}$: $R_\mathrm{sphere}=(3M_\mathrm{cell}/(4\pi\rho_\mathrm{gas}))^{1/3}$, where $M_\mathrm{cell}\approx1.4\times10^6\,\mathrm{M}_\odot$ is the mass of a TNG100 gas cell and $\rho_\mathrm{gas}$ its gas mass density. The \HI column density in \citet{Villaescusa2018} is calculated according to $N_\mathrm{H\,\textsc{i}}=d_\mathrm{segment}\rho_\mathrm{H\,\textsc{i}}/m_\mathrm{H\,\textsc{i}}$, with $d_\mathrm{segment}$ being the segment through the spherically modelled gas cell. Since $d_\mathrm{segment}<2R_\mathrm{sphere}$ and $\rho_\mathrm{HI}<\rho_\mathrm{gas}$, we have $R_\mathrm{sphere}<\sqrt{3M_\mathrm{cell}/(2\pi m_\mathrm{H\,\textsc{i}}N_\mathrm{H\,\textsc{i}})}\approx290\,\mathrm{pc}$ for the TNG100 gas cell mass and $N_\mathrm{H\,\textsc{i}}=10^{21}\,\mathrm{cm}^{-2}$.

For the $20`000\times20`000$ grid used by \citet{Villaescusa2018} to calculate the CDDF, the different line-of-sights are separated by $\approx5\,\mathrm{kpc}$, which means it is very unlikely for the high-density gas to be picked up by the line-of-sights. Since we smooth the Voronoi gas cells when projecting them to a 2D grid we can more accurately take the high-density gas into account. We stress that for our approach for the TNG-intrinsic CDDF (using plain \HI maps), the \HI map resolution (with a default $128\times128$ grid for each galaxy) does not affect the \HI CDDF. The discrepancy in the \HI CDDF between our result and \citet{Villaescusa2018} diminishes at the highest column densities, which we attribute to an underestimate of molecular hydrogen formation in \citet{Villaescusa2018}. This effect increases the \HI CDDF at the highest column densities and hence counters the effect of missing high-density gas cells in \citet{Villaescusa2018}.

Lastly, \citet{Szakacs2022} examined the H$_2$ CDDF in TNG50/TNG100 and its dependency on simulation and map resolution. The H$_2$ fraction is taken from \citet{Popping2019} with different partitioning schemes, and the CDDF is calculated by summing over the contributions from individual galaxies as in the present study. In line with our results, \citet{Szakacs2022} find that the map resolution (changing from 150 pc to 1 kpc maps) hardly affects the CDDF. The simulation resolution (comparing TNG50 and TNG100) affects the H$_2$ CDDF in the very high column density regime ($N_\mathrm{H_2}>10^{22}\,\mathrm{cm}^{-2}$), with more molecular hydrogen formed in TNG100. This contrasts to our finding in Figure~\ref{fig:CDDF_VarSim} where there is more \HI in TNG100 for $N_\mathrm{H\,\textsc{i}}>5\times10^{21}\,\mathrm{cm}^{-2}$ compared to TNG50.

\section{Conclusions}\label{sec:Conclusions}

We postprocessed the IllustrisTNG simulations at redshift zero, partitioning the neutral hydrogen content into its atomic and molecular fractions. We explored how the UV radiation from different stellar populations and the propagation of the UV field through the dusty interstellar medium affect the molecular fractions with the radiative transfer code SKIRT. We generated WHISP-like mock \HI maps for IllustrisTNG galaxies and compared them to 21-cm data from the WHISP survey. To compare these resolved \HI maps, we used non-parametric morphologies and the column density distribution function. Our main findings are summarized as follows:

\begin{itemize}

    \item Realistic UV fields taking dust attenuation into account can substantially affect the \HI distribution for a subset of individual galaxies, but for statistical averages such as the H{\,\sc{i}}/H$_2$ mass functions (Figure~\ref{fig:MassFunction}) and average radial profiles (Figure~\ref{fig:Profile}) the dust attenuation effect is negligible compared to the optically thin (Diemer) scheme.

    \item If the UV radiation is not propagated at all (Lagos scheme), significant statistical differences compared to the SKIRT/Diemer schemes arise. For instance, the \HI mass function is underestimated by 25\,\% at the high-mass end, while the H$_2$ mass function is overestimated by 25\,\% for $M_\mathrm{H_2}>3\times10^8\,\mathrm{M}_\odot$ (Figure~\ref{fig:MassFunction}). We remark that our default partitioning scheme (GD14) minimizes the differences between the various UV fields.

    \item For the 30"-resolution maps (which are most reliable), the non-parametric morphologies of \HI maps of WHISP observational data and mock \HI maps of TNG50 galaxies are in good agreement for the asymmetry, smoothness and Gini statistics (Figure~\ref{fig:MorphologiesVarResWISE}). On the other hand, the TNG50 galaxies feature lower $C$ and higher $M_{20}$ values (Figure~\ref{fig:MorphologyScatter}). Visual inspection of the TNG50 \HI maps reveals that a substantial amount of TNG50 galaxies exhibits large central \HI holes, which are not seen in the WHISP data. The TNG50 galaxies with a very low concentration statistic are exclusively face-on, where the impact of the central \HI hole on $C$ is maximized.

    \item We attribute the prevalence of central \HI holes in TNG50 galaxies mostly to feedback from AGN, which ionizes and/or expels the neutral hydrogen gas from galaxy centers. Excluding galaxies that experienced feedback from the kinetic channel or above-average energy injection from the thermal channel reduces the tension in the $C$ and $M_{20}$ statistics. Switching to the lower-resolution TNG100 simulation (the resolution at which the IllustrisTNG physical model is calibrated) also lowers the tension.

    \item The \HI column density distribution function (CDDF) of TNG50 differs from the one based on WHISP data. TNG50 contains more \HI at intermediate column densities ($N_\mathrm{H\,\textsc{i}}\approx10^{20}-20^{21}\,\mathrm{cm}^{-2}$), but less high-column-density \HI gas compared to WHISP (Figure~\ref{fig:CDDF}). As the bulk of the \HI abundance ($\Omega_\mathrm{H\,\textsc{i}}$) stems from intermediate column densities, TNG50 also has $\Omega_\mathrm{H\,\textsc{i}}$ exceeding observational estimates based on blind \HI surveys (e.g. ALFALFA). The lack of high-column \HI in TNG50 (and IllustrisTNG in general) is due to the beam smoothing in the mock \HI map creation, an effect that was neglected in previous studies of the \HI CDDF in cosmological simulations (\citealt{Rahmati2013a}; \citealt{Rahmati2013b}; \citealt{Villaescusa2018}).

    \item As the bulk of the \HI gas resides in a sweet spot for observational detection according to TNG50 (both in terms of \HI masses and \HI column densities), we attribute the $\Omega_\mathrm{H\,\textsc{i}}$ tension to an overabundance of atomic hydrogen in TNG50 (and not to an underestimation of the observational value). For the lower-resolution TNG100 simulation, the \HI abundance agrees with the observed value, leading to a better match in the CDDF as well. This result is expected as a TNG50 galaxy will generally have higher stellar and gas masses compared to a TNG100 galaxy in a similar dark matter halo, due to the IllustrisTNG model being calibrated at the TNG100 resolution.

    \item The mismatch in the CDDF at the high-column-density end can be remedied by using local partitioning schemes (based on number instead of column densities). Local partitioning schemes also increase the atomic fractions in galaxy centers, which increases the concentration statistic of TNG50 \HI maps and reduces the tension with the WHISP morphologies. Hence, we hypothesize that the conventional column-based partitioning schemes fail at the highest column densities as the Jeans approximation could break down. The drawback of the local partitioning schemes is that they significantly overpredict the overall abundance of atomic hydrogen. Hence, albeit conceptually appealing, we refrain from using local partitioning schemes for the main results of this paper.

\end{itemize}

Based on the various partitioning schemes and UV fields explored in this study, we recommend to use the column-based (not the local) GD14 model for hydrogen partitioning in cosmological simulations. The GD14 model is an update of the GK11 model, contains a factor to take the cell size into account to mitigate resolution dependencies, and nicely couples conceptually to cosmological simulations as it is itself a simulation-calibrated partitioning scheme. Furthermore, it predicts the largest H$_2$ fractions of all partitioning schemes considered in this study (except for S14, but this partitioning scheme leads to the most significant tensions in the non-parametric morphologies and the CDDF comparing TNG50 and WHISP), such that it provides the best match to the observational value of $\Omega_\mathrm{H\,\textsc{i}}$.

For the UV fields, we recommend to always spread the UV flux, i.e. to use the Diemer or SKIRT schemes. The effect of dust attenuation is negligible for statistical properties of the galaxy population like the \HI CDDF or the HIMF. However, the dust attenuation effect can be significant for individual galaxies and becomes more pronounced for molecular hydrogen properties. Studies that consider H$_2$ in smaller galaxy samples or zoom-simulations could obtain significantly biased results if not correcting for the effect of dust attenuation.

\section*{Acknowledgments}

We wish to express our gratitude towards Ana Trčka who helped setting up the SKIRT analysis of the present study. We also wish to thank Benne Holwerda and Nick Gnedin for fruitful discussions about non-parametric morphologies and partitioning schemes, and are grateful for feedback from Annalisa Pillepich and Matthew Smith.

AG gratefully acknowledges financial support from the Fund for Scientific Research Flanders (FWO-Vlaanderen, project FWO.3F0.2021.0030.01). DN acknowledges funding from the Deutsche Forschungsgemeinschaft (DFG) through an Emmy Noether Research Group (grant number NE 2441/1-1). This work has received funding from the European Research Council (ERC) under the European Union’s Horizon 2020 research and innovation programme (grant agreement No 882793 `MeerGas').

This study made extensive use of the \texttt{Python} programming language, especially the \texttt{numpy} (\citealt{numpy}), \texttt{matplotlib} (\citealt{matplotlib}), and \texttt{scipy} (\citealt{scipy}) packages.

The WHISP observation were carried out with the Westerbork Synthesis Radio Telescope, which is operated by the Netherlands Foundation for Research in Astronomy (ASTRON) with financial support from the Netherlands Foundation for Scientific Research (NWO). The WHISP project was carried out at the Kapteyn Astronomical Institute by J. Kamphuis, D. Sijbring and Y. Tang under the supervision of T.S. van Albada, J.M. van der Hulst and R. Sancisi.

\section*{Data and code availability}

The IllustrisTNG data used in this work is publicly available at \url{https://www.tng-project.org/} as described by \citet{Nelson2019dataRelease}. The WHISP moment-zero \HI maps in all resolutions are publicly available at \url{http://wow.astron.nl/}. WISE fluxes for the WHISP galaxies are publicly available at \url{https://academic.oup.com/mnras/article/502/4/5711/6095720} (\citealt{Naluminsa2021}). Distances to WHISP galaxies are taken from the publicly available NED database (\url{https://ned.ipac.caltech.edu/}).

The SKIRT code (version 9) is publicly available at \url{https://skirt.ugent.be/root/_home.html} and described by \citet{Camps2020}. The \texttt{statmorph} code (\citealt{Rodriguez-Gomez2019}) is publicly available at its github repository (\url{https://github.com/vrodgom/statmorph}). The algorithm to spread the UV field in an optically thin fashion (Diemer UV field) as well as the projection algorithm is part of the private \texttt{hydrotools} repository of Benedikt Diemer. We are happy to share all other parts of the code and generated data of this work upon request.

%%%%%%%%%%%%%%%%%%%%%%%%%%%%%%%%%%%%%%%%%%%%%%%%%%



\bibliography{bibliography}{}
\bibliographystyle{mnras}

%%%%%%%%%%%%%%%%%%%%%%%%%%%%%%%%%%%%%%%%%%%%%%%%%%

%%%%%%%%%%%%%%%%% APPENDICES %%%%%%%%%%%%%%%%%%%%%

\appendix

\section{Implementation of Partitioning Schemes}

\subsection{Krumholz, McKee \& Tumlinson 2009 (KMT09)}\label{sec:KMT}
\citet{Krumholz2009} (KMT09) consider an analytical model of molecular hydrogen formation and dissociation akin to S14. The model is equivalent to K13 except for the estimation of $\chi$ which quantifies the balance between UV photodissociation and recombination of H$_2$. While this parameter depends on $U_\mathrm{MW}$ in \citet{Krumholz2013}, it is estimated from the metal mass fraction in the gas cell, $Z$ (Eqn. 7 in \citealt{Krumholz2009} or Eqn.4 in \citealt{Krumholz2011}):

\begin{equation}
    \chi=3.1\Biggl(\frac{1+3.1(Z/Z_\odot)^{0.365}}{4.1}\Biggr).
\end{equation}
The molecular hydrogen fraction as well as the $s$ and $\tau_c$ parameter are identically calculated as in the K13 model described in appendix C.4 of \citet{Diemer2018} (their Eqns. 18, 19 and 26). We remark that this partitioning scheme is used to partition the hydrogen in the cosmological simulations MUFASA (\citealt{Dave2016}) and Simba (\citealt{Dave2019}), but with a clumping factor of $\approx10$ and the Sobolev approximation when calculating $\tau_c$ (we use a clumping factor of five and the Jeans approximation following \citealt{Diemer2018}).

\subsection{Gnedin \& Kravtsov 2011 (local version)}\label{sec:GK11 local}

This partitioning scheme is the local version (having number instead of column densities as input) of the GK11 scheme. We refer to appendix C2 of \citet{Diemer2018} for the description of the $g$ parameter (their equations 11 and 12), and write only the new equations for this local, UV-dependent partitioning scheme:

\begin{equation}
    f_\mathrm{mol}=\frac{1}{1+\exp(-4x-3x^3)},
\end{equation}

\begin{equation}
    x=\Lambda^{3/7}\ln\Bigl(\frac{D_\mathrm{MW}n_\mathrm{H}}{\Lambda n_\ast}\Bigr),
\end{equation}
where $n_\mathrm{H}$ is the total hydrogen number density in protons per $\mathrm{cm}^3$, $n_\ast=25\,\mathrm{cm}^{-3}$, and $\Lambda$ is

\begin{equation}
    \Lambda=\ln\Bigl(1+gD_\mathrm{MW}^{3/7}(U_\mathrm{MW}/15)^{4/7}\Bigr).
\end{equation}
Note that $f_\mathrm{mol}$ in this formalism technically corresponds to the mass fraction in molecular hydrogen compared to total hydrogen, while we use $f_\mathrm{mol}$ for the mass fraction in H$_2$ compared to \textit{neutral} hydrogen. We use the two quantities interchangeably as under all conditions of interest, the ionized hydrogen fraction is negligible when molecular hydrogen forms.

\subsection{Gnedin \& Draine 2014 (local version)}\label{sec:GD14 local}

Equivalently, this is the local version of the GD14 partitioning scheme. As this varies quite substantially from their partitioning scheme based on surface densities presented in appendix C3 of \citet{Diemer2018} we note the full set of equations here:

\begin{equation}
    f_\mathrm{mol}=\frac{1}{1+\exp\bigl(-x(1-0.02x+0.001x^2)\bigr)},
\end{equation}

\begin{equation}
    x=w\ln\Bigl(\frac{n_\mathrm{H}}{n_{1/2}}\Bigr),
\end{equation}
where $n_{1/2}$ is the hydrogen number density at which $f_\mathrm{mol}=0.5$.

\begin{equation}
    n_{1/2}=\frac{n_\ast\Lambda}{g},
\end{equation}

\begin{equation}
    w=0.8+\frac{\Lambda^{1/2}}{S^{1/3}},
\end{equation}

\begin{equation}
    \Lambda=\ln\Bigl(1+(0.05/g+U_\mathrm{MW})^{2/3}g^{1/3}/U_\ast\Bigr),
\end{equation}

\begin{equation}
    n_\ast=14\,\mathrm{cm}^{-3}\frac{D_\ast^{1/2}}{S},
\end{equation}

\begin{equation}
    U_\ast=\frac{9D_\ast}{S},
\end{equation}

\begin{equation}
    D_\ast=0.17\frac{2+S^5}{1+S^5},
\end{equation}
and $S=L_\mathrm{cell}/100\,\mathrm{pc}$ is a factor to correct for the spatial resolution of the simulation. Following \citet{Diemer2018} we approximate $L_\mathrm{cell}=(m/\rho)^{1/3}$ for each gas cell. We remark that \citet{Gnedin2014} calibrated this resolution correction up to cell sizes of 500 pc, which should be sufficient to reach the TNG50 cell sizes at which molecular hydrogen formation becomes important.

\section{Projected modelling}\label{sec:Projected modelling}

\begin{figure*}
    \centering
    \includegraphics[width=\textwidth]{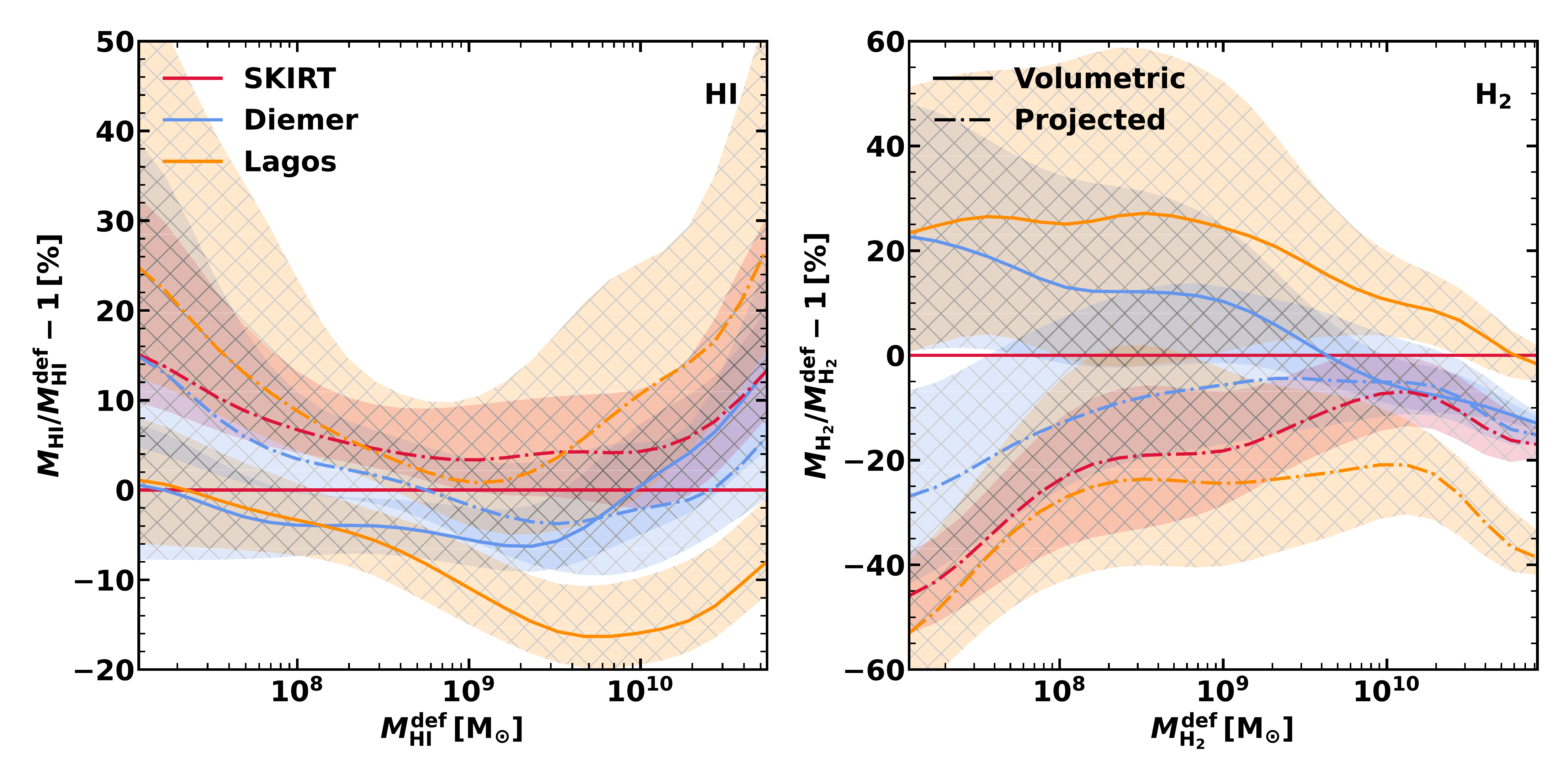}
    \caption{Mass differences between the different UV field estimates and volumetric/projected modelling for TNG50. We only use the default partitioning scheme of GD14 here. Shown is the difference to the default volumetric SKIRT model (`def'). The lines and hatched areas indicate the running medians and interquartile ranges for the individual H{\,\sc{i}}/H$_2$ masses of the galaxy population as a function of their H{\,\sc{i}}/H$_2$ mass of the default model. For a better visualization we smoothed all curves with a Gaussian filter with $\sigma\approx0.23\,\mathrm{dex}$.}
    \label{fig:MassDiff}
\end{figure*}

A troublesome aspect of hydrogen partitioning in cosmological simulations is the usage of surface (or column) densities in partitioning schemes, which are difficult and somewhat ambiguous to extract from the simulations. The approach adopted here and in most other studies is to calculate surface densities of gas cells with the Jeans approximation (`volumetric modelling', Eqn.~\ref{eq:Jeans}). \citet{Diemer2018} find that the Jeans approximation introduces significant biases in the estimated surface densities. A possible resolution to overcome this issue is the usage of local partitioning schemes as discussed in Section~\ref{sec:LocalSchemes}. Another alternative suggested by \citet{Diemer2018} is to estimate surface densities by rotating galaxies into a face-on position, projecting the gas cells on a 2D grid to compute surface densities, and performing the hydrogen partitioning directly on this 2D grid (`projected modelling').

For the analytical and simulation-calibrated partitioning schemes considered here, \citet{Diemer2018} find that the volumetric and projected approaches lead to similar results on average for the entire galaxy population, but with large scatter for individual galaxies (see their figure 4). It is not obvious which approach should be preferred, as surface densities have different meanings across the various partitioning schemes. For instance, S14 consider an idealized interstellar gas cloud as a 1D slab, while GK11 consider columns calculated from lower-resolution gas cells with a size of $\approx500\,\mathrm{pc}$ corresponding to the average thickness of galactic disks in their simulations.

An advantage of projected modelling is that it is significantly faster than volumetric modelling. For our purposes, however, an important shortcoming of the projected approach is that one can only generate maps in face-on projection. While this is fine to calculate masses and radial profiles, for the mock \HI maps that we compare to the WHISP survey (which contains galaxies in random orientation) we cannot use the projected approach there.

We test here if the finding of \citet{Diemer2018} that the volumetric and projected approaches agree on average (for each partitioning scheme) extends to the SKIRT and Lagos UV fields. To this end we calculate \HI and H$_2$ masses both in the volumetric and projected incarnation with our default partitioning scheme for the base TNG50 sample. For the projected modelling we generate the face-on maps of all quantities required for the hydrogen partitioning using the plain map algorithm (Section~\ref{sec:PlainMaps}).

Figure~\ref{fig:MassDiff} shows the deviation in galaxy H{\,\sc{i}}/H$_2$ mass from the default \HI model (volumetric modelling with the SKIRT UV field) when changing the UV field and/or switching to projected modelling. The lines indicate the running medians, the hatched areas correspond to the interquartile range. Unsurprisingly, the Lagos models deviate most from the default model, furthermore they also feature the largest difference between the volumetric and projected version (solid and dashed orange lines). For intermediate H$_2$ masses of $\sim10^9\,\mathrm{M}_\odot$, projected modelling leads to H$_2$ masses lower by a factor of two compared to volumetric modelling (with the Lagos UV field). 

For the SKIRT and Diemer UV fields the discrepancies between the volumetric and projected models are mostly not significant as the interquartile ranges overlap. For our estimate of $\Omega_\mathrm{H\,\textsc{i}}$ with the default \HI model (volumetric SKIRT model), switching to the projected SKIRT model would slightly increase the \HI abundance in IllustrisTNG by $\approx5\,\%$. Hence, this systematic uncertainty arising from volumetric/projected modelling does not affect the conclusions from Section~\ref{sec:HIabundance}. We note that the default model seems to underestimate the largest \HI masses by $\sim20\,\%$ compared to all other five models. Furthermore, we remark that although the mass functions between the different UV field estimates are mostly consistent (Figure~\ref{fig:MassFunction}), individual galactic H{\,\sc{i}}/H$_2$ masses can differ a lot between different models, especially when comparing the Lagos UV field with the SKIRT/Diemer UV fields.

%%%%%%%%%%%%%%%%%%%%%%%%%%%%%%%%%%%%%%%%%%%%%%%%%%

% Don't change these lines
\bsp	% typesetting comment
\label{lastpage}
\end{document}